\documentclass[a4paper,fleqn,usenatbib]{mnras}

\usepackage{mathptmx}
\usepackage{txfonts}
\hypersetup{draft}

\usepackage[T1]{fontenc}
\usepackage{ae,aecompl}

\usepackage{graphicx}   
\usepackage{amssymb}    

\title[67P dust properties]{Comet 67P outbursts and quiescent coma at 1.3 AU from the Sun: dust properties from Rosetta/VIRTIS-H observations}

\author[D. Bockel\'ee-Morvan et al.]{Dominique Bockel\'ee-Morvan,$^{1}$\thanks{E-mail: dominique.bockelee@obspm.fr}
G. Rinaldi,$^{2}$
S. Erard,$^{1}$
C. Leyrat,$^{1}$
F. Capaccioni,$^{2}$
\newauthor
P. Drossart,$^{1}$ G. Filacchione,$^{2}$  A. Migliorini,$^{2}$ E.
Quirico,$^{3}$ S. Mottola,$^{4}$
 G. Tozzi$^{5}$
\newauthor
G. Arnold,$^{4}$ N. Biver,$^{1}$ M. Combes,$^{1}$\thanks{deceased}
J. Crovisier,$^{1}$
 A. Longobardo,$^{2}$ M. Blecka,$^{6}$  
 \newauthor
 M.-T. Capria,$^{2}$
\\
$^{1}$ LESIA, Observatoire de Paris, PSL Research University, CNRS, Sorbonne Universit\'es, UPMC Univ. Paris 06, \\
Univ. Paris-Diderot, Sorbonne Paris Cit\'e, 5 place Jules Janssen, 92195 Meudon, France\\
$^{2}$ INAF-IAPS, Istituto di Astrofisica e Planetologia Spaziali, via del fosso del Cavaliere, 100, 00133, Rome, Italy \\
$^{3}$ Universit\'e Grenoble Alpes, CNRS, Institut de Plan\'etologie et d'Astrophysique de Grenoble, \\ 414 rue de la Piscine, BP53, 38041, Grenoble, France  \\
$^{4}$ Institute for Planetary Research, Deutsches Zentrum f\"{u}r Luft- und Raumfahrt (DLR), Rutherfordstrasse 2, 12489, Berlin, Germany \\
$^{5}$ INAF, Osservatorio Astrofisico di Arcetri,  Largo E. Fermi
5, 50125 Firenze, Italy\\ $^{6}$ Space Research Centre, Polish
Academy of Sciences, Bartycka 18a, 00716, 
Warsaw, Poland
}

\date{Accepted XXX. Received YYY; in original form ZZZ}

\pubyear{2016}

\begin{document}
\label{firstpage}
\pagerange{\pageref{firstpage}--\pageref{lastpage}}
\maketitle

\begin{abstract}
    We present 2--5 $\mu$m spectroscopic observations of the dust coma of 67P/Churyumov-Gerasimenko
    obtained with the VIRTIS-H instrument onboard Rosetta during two
outbursts that occurred on 2015, 13 September 13.6 h UT and 14
September 18.8 h UT at 1.3 AU from the Sun.  Scattering and thermal
properties measured before the outburst are in the mean of values
measured for  moderately active comets. The colour
temperature excess (or superheat factor) can be attributed to submicrometre-sized  particles composed of absorbing material or to porous fractal-like
aggregates such as those collected by the Rosetta in situ dust
instruments. The power law index of the dust size distribution is
in the range 2--3.  The sudden increase of infrared emission
associated to the outbursts is correlated with a large increase of
the colour temperature (from 300 K to up to 630 K) and a change of
the dust colour at 2--2.5 $\mu$m from red to blue colours,
revealing the presence of very small grains ($\leq$ 100 nm) in the
outburst material. In addition, the measured large bolometric
albedos ($\sim$ 0.7) indicate bright grains in the ejecta, which
could either be silicatic grains, implying the thermal degradation
of the carbonaceous material, or icy grains. The 3-$\mu$m
absorption band from water ice is not detected in the spectra
acquired during the outbursts, whereas signatures of organic
compounds  near 3.4 $\mu$m are observed in emission. The H$_2$O
2.7-$\mu$m and CO$_2$ 4.3-$\mu$m vibrational bands do not show any
enhancement during the outbursts.


\end{abstract}



\section{Introduction}

The Rosetta mission of the European Space Agency accompanied comet 67P/Churyumov-Gerasimenko between 2014 and 2016 as it approached perihelion (13 August 2015) and receded from the Sun. The suite of instruments on the orbiter and on the Philae lander
provided complementary information on the physico-chemical properties of the nucleus, and inner coma gases and dust, from which questions related to the formation of 67P and origin of its primary constituents could be studied \citep{Davidsson2016,Fulle2016}. Rosetta also provided the first opportunity to study closely how cometary activity proceeds, evolves with seasons, and affects the nucleus surface properties \citep{El-Maary2017}.

One of the striking discoveries of Rosetta has been the repeatibility of jets from one rotation to another, growing and fading according to solar illumination, and originating from cliffs, fractures and pits \citep{Vincent2016a}. Strong similarities  between the near-nucleus dust and water distributions were observed, indicating that dust release is associated to water sublimation \citep{Rinaldi2016,Fink2016}. Other manifestations of cometary activity are the detection of morning jets \citep{Shi2016} and transient events, the so-called outbursts. The cameras of the Optical, Spectroscopic, and Infrared Remote Imaging System (OSIRIS) detected not less than 34 outbursts during the three months encompassing the perihelion passage, one every 2.4 rotations, with three main plume morphologies: a narrow jet, a broad fan, and more complex plumes featuring both previous types together \citep{Vincent2016b}. \citet{Gruen2016} review observations by nine Rosetta instruments of a dusty outburst which occurred on 19 February 2016, when the comet was at 2.4 AU from the Sun. Gaseous outbursts with no dust counterparts were also observed \citep{Feldman2016}.

A few outbursts have been observed with the Visible InfraRed Thermal Imaging
Spectrometer (VIRTIS) onboard Rosetta \citep{Coradini2007}, in both VIRTIS-H and VIRTIS-M channels. This paper presents the analysis of VIRTIS-H infrared  (2--5 $\mu$m) spectra  acquired during two dusty outbursts that occurred on 13 and 14 September 2015, i.e., about one month after perihelion.
The complementary spectro-imaging observations of these outbursts undertaken with VIRTIS-M in the visible spectral range are presented by \citet{Rinaldi2017}. The two outbursts have very different morphologies, with the 13 September outburst being strongly radially collimated, whereas the 14 September outburst extended laterally \citep{Rinaldi2017}. The VIRTIS-H spectra cover a range where both the dust-scattered sunlight and the dust thermal emission can be studied, to derive constraints on the properties of the dust particles present in the ejecta cloud. The 2--5 $\mu$m spectral range also covers vibrational bands of important gases, such as H$_2$O and CO$_2$, so that the gaseous counterpart of the outbursts can be potentially studied.

Section \ref{sec:obs} presents the observations. In Sect.~\ref{sec:prop}, the methodology to reduce and analyse the VIRTIS-H spectra, and the retrieved spectral properties are given. Section \ref{sec:analysis} presents the models used to interpret the spectra, and the constraints derived on the dust size distribution and composition. Attention is given to the properties of the dust coma both before the outburst and in the outburst ejecta. The gaseous counterpart is discussed in Sect.~\ref{sec:gas}.
A discussion follows in Sect.\ref{sec:discussion}.


\section{Outburst observations}
\label{sec:obs}
VIRTIS obtained spectra of the coma of comet 67P at the
time of two important outbursts on 13 and 14 September 2015. VIRTIS is composed of two
channels: VIRTIS-M, a spectro-imager operating both in the visible
(0.25--1 $\mu$m) and infrared  (1--5 $\mu$m) ranges at moderate
spectral resolution ($\lambda$/$\Delta \lambda$ = 70-380), and
VIRTIS-H, a cross-dispersing spectrometer providing spectra with higher spectral resolution capabilities
($\lambda$/$\Delta \lambda$ = 1300-3500) in eight
orders of a grating covering the range 1.9--5.0 $\mu$m \citep{Drossart2000,Coradini2007}.


As for most Rosetta instruments, the line of sight of VIRTIS-H is along the Z-axis of the spacecraft (S/C). The instantaneous field of view (FOV) of this
point instrument is 0.58 $\times$ 1.74 mrad$^2$ (the larger
dimension being along the Y axis). Details on the calibration process are given in \citet{dbm2016}.
The version of the calibration pipeline is CALIBROS--1.2--150126.

\begin{figure}
    \includegraphics[width=\columnwidth]{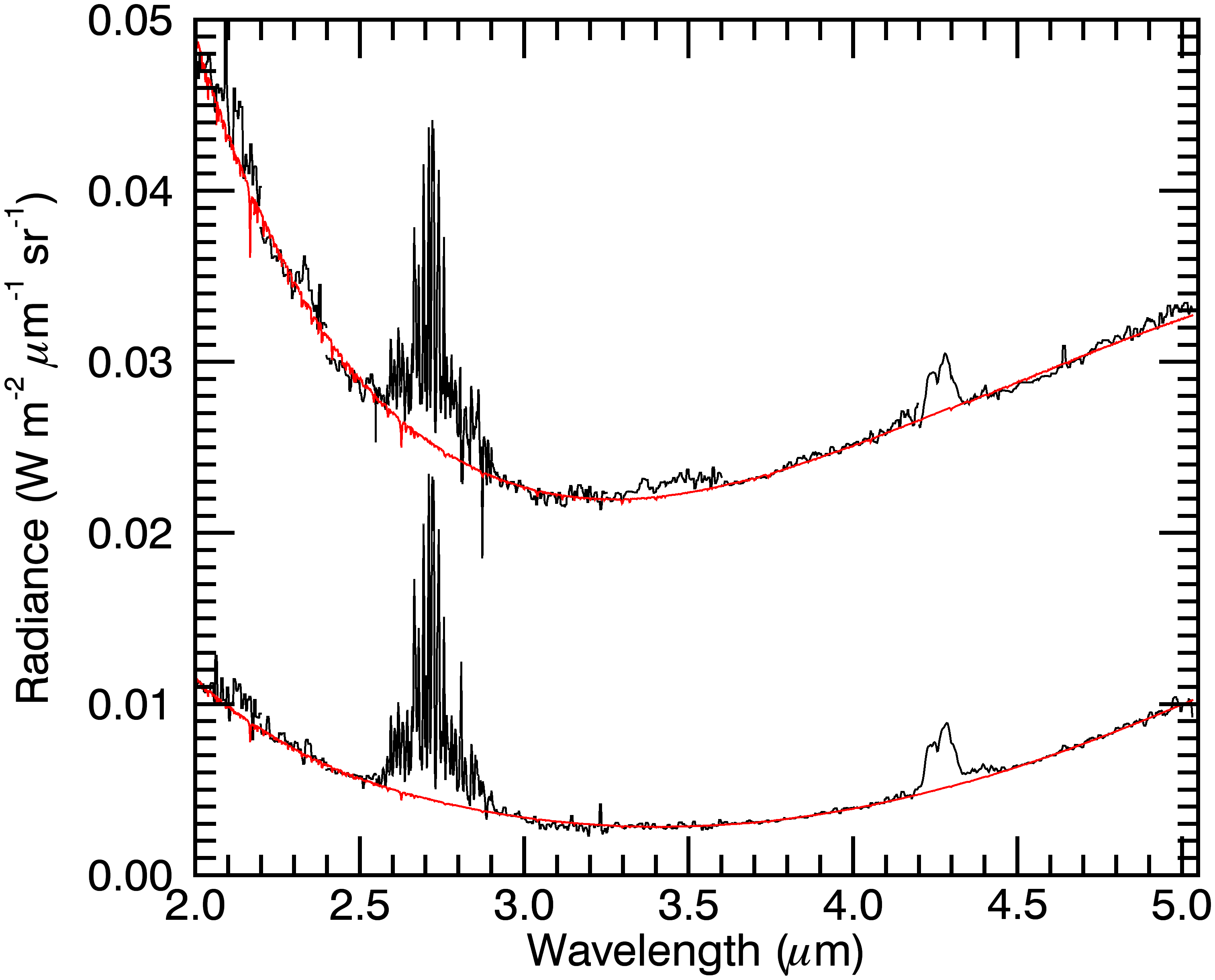}
    \includegraphics[width=\columnwidth]{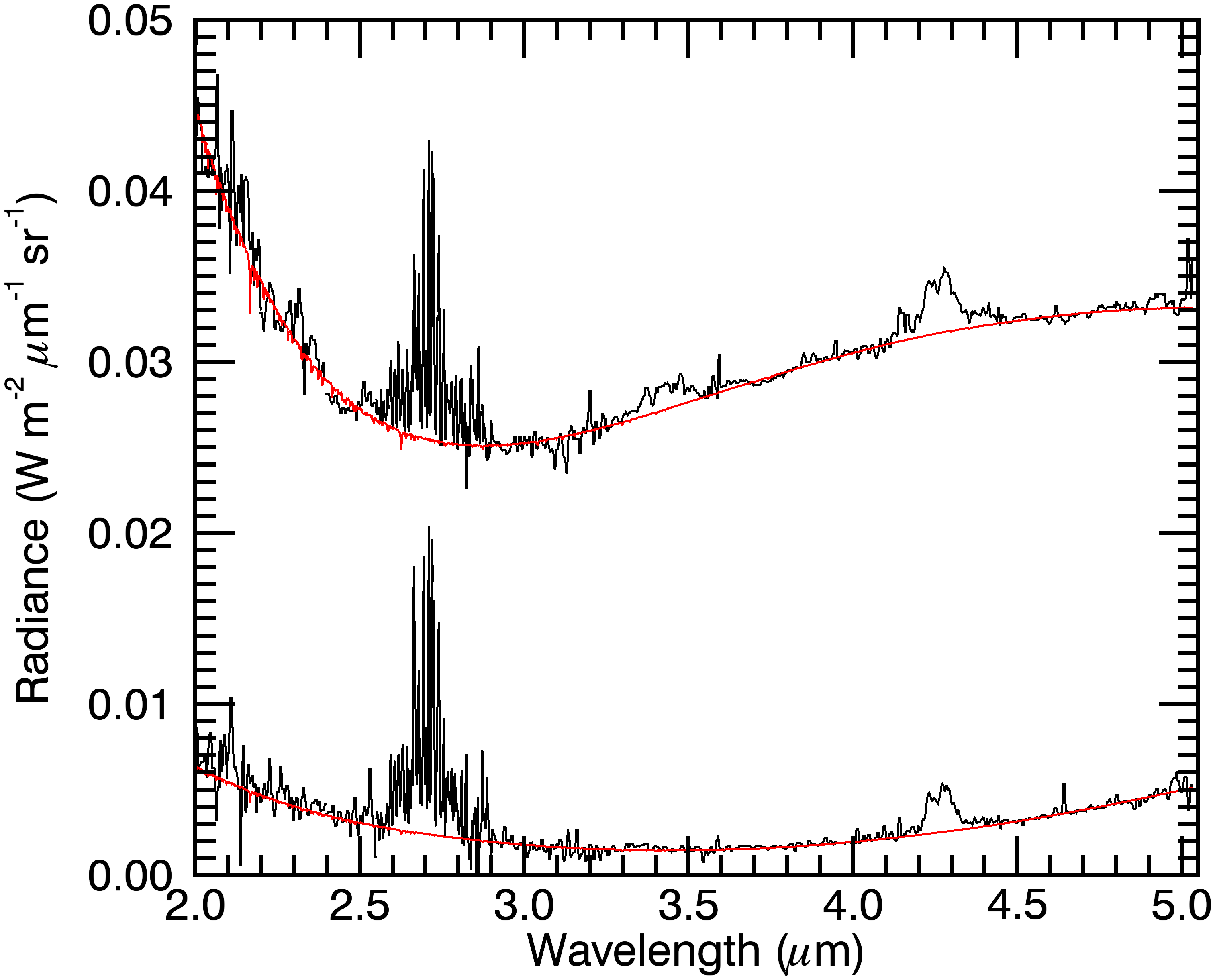}
       \caption{VIRTIS-H spectra acquired on 13 September (top) and 14 September (bottom) 2015. In each plot, the upper and lower spectrum were acquired during the outburst and before the ourburst, respectively. For 13  Sept., the time intervals are 13.4003--13.6028 h and 13.7125--13.7968 h UT. For 14 Sept., the time intervals are 18.8060--18.8248 h and 18.8632--18.8838 h UT. Outburst spectra were shifted by +0.01 W m$^{-2}$ $\mu$m$^{-1}$ sr$^{-1}$. Median smoothing (with dimension 20 outside the molecular bands) was applied. Radiance values in the range 4.2--5 $\mu$m were corrected as described in Sect.~\ref{sec:reduction}, with $s$=0.2. Overplotted, in red, is the best model fit with the  retrieved parameters given in Table~\ref{tab:properties}.      }
    \label{fig:spectra}
\end{figure}

\begin{figure}
    \includegraphics[width=\columnwidth]{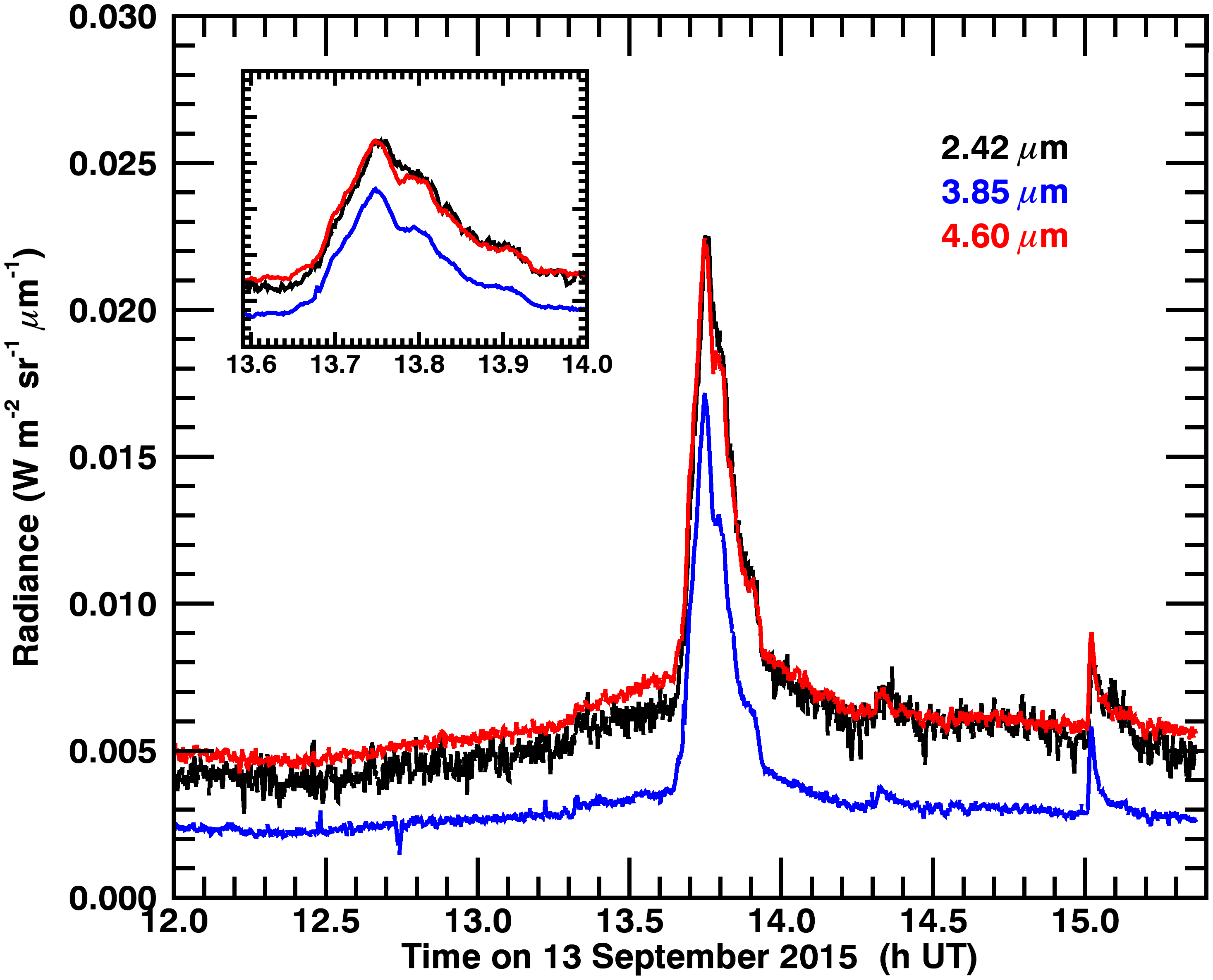}
    \includegraphics[width=\columnwidth]{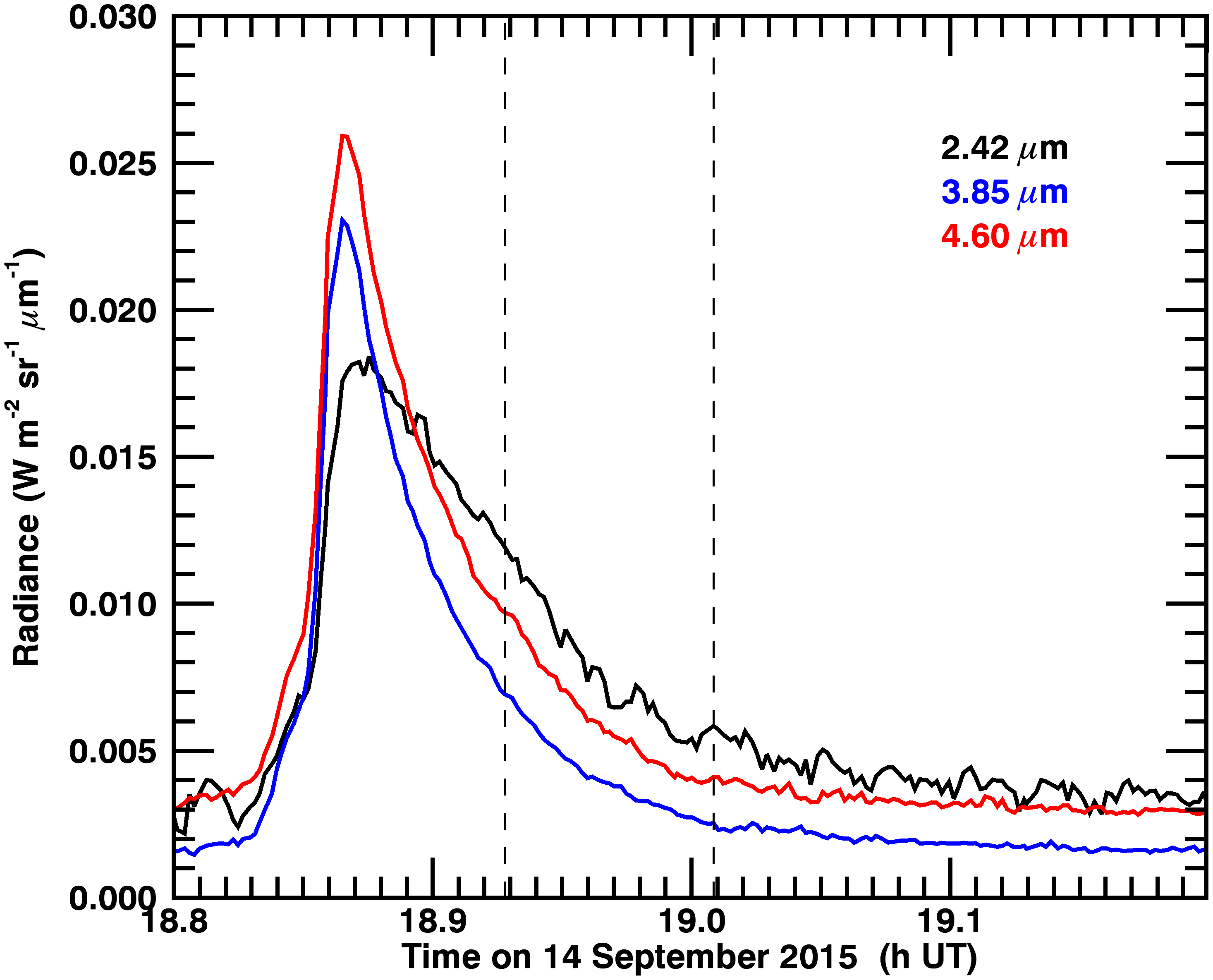}
       \caption{Time evolution of the radiance at 2.46, 3.85, and 4.60 $\mu$m on 13 September (top) and 14 September (bottom) 2015. The dashed lines in the bottom figure delimit the period when Rosetta was slewing. Outside this time range the FOV was staring at a fixed position (see text).  }
    \label{fig:lightcurve}
\end{figure}

The 13 and 14 September outbursts were first identified in images obtained by VIRTIS-M at 0.55 $\mu$m, showing unusual coma morphologies \citep{Rinaldi2017}. In VIRTIS-H data, these outbursts are characterised by a sudden increase of the dust continuum, both in reflected light ($<$ 3 $\mu$m) and thermal emission ($>$ 3 $\mu$m) (Figs~\ref{fig:spectra}-\ref{fig:lightcurve}), followed by a gentle decrease lasting a few minutes to tens of  minutes. Table~\ref{tab:log} provides
some details on the two VIRTIS-H observations that are considered: identification number of the data cube (Obs. Id), start time and duration of the observations, number of acquisitions and acquisition duration. For these observations, the exposure time was 3 s, and the dark rate was 4 (one dark image every 4 acquisitions). Frame summation was performed on-board, by summing pairs of acquisitions, so that the duration of each acquisition in the calibrated data cubes is 6 s.

Geometric parameters are provided in Table~\ref{tab:log}: the distance of Rosetta to the comet $\Delta$(S/C), the heliocentric distance $r_h$, the phase angle, and some information on the position of the FOV with respect to the comet and comet-Sun line, that is its distance $\rho$ to comet centre, and the position angle $PA$ in
the XY plane. In our convention \citep[see][]{dbm2016}, the Sun is at $PA$=270$^{\circ}$, which means that
the FOV during the measurements ($PA$=265$^{\circ}$, Table~\ref{tab:log}) was almost along the comet-Sun line. For the 13 September data cube,
the FOV was stared at a fixed distance in inertial frame ($\rho$ = 3.61 km). On the other hand, for the 14 September data cube, several (actually 12) stared positions were commanded along the -X direction (i.e., towards Sun), moving back and forth from 4.82 to 9.32 km from comet centre. Fortunately, the recorded 14 September outburst occurred just after the beginning of a stared sequence at $\rho$ = 4.835 km which started at 18.804164 h UT and ended at 18.927901 h UT. The next stared sequence at $\rho$ = 6.319 km started at 19.008519 h UT, when some outburst material was still crossing the FOV (Fig.~\ref{fig:lightcurve}). Between the two sequences, the Rosetta spacecraft was slewing and the FOV was at intermediate distances.

The time, duration, and strength of the outbursts are given in
Table~\ref{tab:properties}. The duration is computed using the time at which the radiance returned at the pre-outburst value. The 13 September main outburst
started at 13.645 UT, and was also observed with the Alice UV
spectrometer (Feldman, private communication).  The time evolution
of the radiance then followed a complex pattern suggesting
consecutive events (at least three, Fig.~\ref{fig:lightcurve}).
This outburst, which lasted about 30 min, was followed by two
"mini outbursts", which occurred about 40 min and 82 min later.
The 14 September outburst started at 18.828 UT, and can be
qualified as a major outburst since the dust thermal emission at
4.6 $\mu$m increased by a factor of $\sim$7 with respect to the
pre-outburst value. All the outbursts are characterised by a
sudden increase of the dust emission, peaking a few minutes later,
followed by a smooth decrease of the emission until it comes back
to the pre-outburst value. Another common feature is the change in
the ratio of the intensities at 3.6 and 3.85 $\mu$m. The
pre-outburst values of $I_{\rm 4.6\mu m}$/$I_{\rm 3.85\mu m}$ are
$\sim$ 2. When outburst material is within the FOV, this ratio is
lower, reaching values of 1.4 and 1.7 for the 13 and 14 September
main outbursts, respectively, and 1.9 for the last mini outburst.
This indicates that the particles emitted during the outbursts
have properties that are different from the usual dust. In
particular, this trend would be consistent with higher grain
temperatures in the outburst ejecta. The different lightcurves
observed in reflected and thermal emission on 14 September (Fig.~
\ref{fig:lightcurve}) are another indication of different dust
properties that changed with time.

\begin{table*}
    \caption{Log of VIRTIS-H observations.}
    \label{tab:log}
    \begin{tabular}{lcccccclcc}
        \hline        \noalign{\smallskip}
Obs Id & Start time & Obs length & Nb acq. & Acq. length & $r_h$ & $\Delta$(S/C)   & phase & \multicolumn{2}{c}{Pointing$^b$} \\
&& &&&&&& $\rho$& $PA$ \\
& (UT) & (h)&  & (s) & (AU) & (km)  & ($^{\circ}$) & (km) &  ($^{\circ}$) \\
        \hline
        \noalign{\smallskip}
        00400765976 & 2015-09-13T11:46:19.3 &  3.64  & 1728 & 6 & 1.302 &   312 &   108
&  3.61 &  265  \\
  00400873041 & 2015-09-14T17:30:48.1 & 3.91 & 1856 & 6 & 1.306 &   315 &    99
&  4.82--9.32 &  265 \\
        \hline\noalign{\smallskip}
    \end{tabular}

{\raggedright
    $^a$ This is a factor of two higher than the exposure time of 3 s, because of onboard frame summing.

    $^b$ Distance to comet centre ($\rho$) and position angle $PA$ of the FOV ($PA$(Sun) = 270$^{\circ}$).  During 14-September cube acquisition, the FOV was stared at four different distances in the range indicated.

 }
\end{table*}

\begin{table*}
    \caption{Properties of outburst and pre-outburst spectra.}
    \label{tab:properties}
    \begin{tabular}{llccccccc}
        \hline\noalign{\smallskip}
        Date & Init/peak time$^{(a)}$ & Duration$^{(b)}$ & $I^{\rm max}$/$I^{\rm pre (c)}$ & $T_{\rm col}$  & $S_{\rm heat}$ & $A(\theta)$ & $S'_{\rm col}$ \\
        & (h) & (min) & & (K) & & & \%/100nm  \\
        \hline\noalign{\smallskip}
        {\it Total spectra} &&&&&&\\
        13 Sept. main outburst & 13.645/13.748 & 26.6 & 3.0 & 429$\pm$9$^d$ & 1.76$\pm$0.04$^d$  & 0.51$\pm$0.02$^d$ & 1.0$\pm$0.3$^d$ \\
        13 Sept. mini-outburst 1 & 14.308/--&  6.8 & 1.1 & -- & -- & -- & -- \\
        13 Sept. mini-outburst 2 & 15.007/15.021 & 9.6  & 1.5 & 356$\pm$9$^e$ & 1.46$\pm$0.04$^e$ & 0.29$\pm$0.09$^e$& 2.2$\pm$0.4 $^e$ \\
        14 Sept. outburst& 18.828/18.865 & 12.2 & 7.3 & 561$\pm$12$^f$  & 2.31$\pm$0.05$^f$ & 0.54$\pm$0.01$^f$ & -2.6 $\pm$1.0$^f$   \\
        \hline\noalign{\smallskip}
        {\it Outburst material} &&&&&&\\
        13 Sept. main outburst &--&--&--& 528$\pm$10$^d$ & 2.17$\pm$0.04$^d$ &  0.67$\pm$0.01$^d$ & --2.1$\pm$0.1$^d$  \\
        13 Sept. mini-outburst 2 &--&--&--& 549$\pm$12$^e$ & 2.25$\pm$0.05$^e$ &  0.62$\pm$0.04$^e$ & --2.2$\pm$0.7$^e$  \\
        14 Sept.  outburst &--&--&--& 621$\pm$12$^f$ & 2.55$\pm$0.05$^f$ &  0.56$\pm$0.01$^f$ & --9.1$\pm$1.4$^f$  \\
        \hline\noalign{\smallskip}
        13 Sept. pre-outburst & -- & -- & -- & 297$\pm$6$^g$ & 1.22$\pm$0.02$^g$ & 0.13$\pm$0.02$^g$  & 2.6$\pm$0.3$^g$    \\
        14 Sept. pre-outburst & -- & -- & -- &  293$\pm$8$^h$ & 1.20$\pm$0.03$^h$ & 0.13$\pm$0.02$^h$  & 2.3$\pm$0.4$^h$  \\
        \hline        \noalign{\smallskip}
    \end{tabular}

{\raggedright
    $^a$ Measured at 4.6 $\mu$m (same values for 3.85 $\mu$m).

    $^b$ Defined by the time at which the intensity returned at the pre-outburst value.

    $^c$ Measured at 4.6 $\mu$m.

    $^d$ From spectrum averaging acquisitions 921--961 (13.7125--13.7968 h UT).

    $^e$ Acquisitions 1541--1551 (15.0220--15.0426 h UT).

    $^f$ Acquisitions 641--646 (18.8632--18.8735 h UT).

    $^g$ Acquisitions 773--869 (13.4003--13.6028 h UT).

    $^h$ Acquisitions 614--623 (18.8060--18.8248 h UT).

    }
\end{table*}

\section{Properties of dust ejecta}
\label{sec:prop}

\subsection{Approach for dust spectral properties retrievals}
\label{sec:approach}

\citet{Gehrz1992} described two empirical parameters derivable from the dust infrared spectral energy distribution (SED) measured in cometary atmospheres that are related to the optical properties of the grains contributing
importantly to the SED. These parameters are the superheat $S_{\rm heat}$ and the bolometric phase-dependent albedo $A$($\theta$), where $\theta$ is the scattering angle.

The superheat is defined as the ratio of the observed colour temperature $T_{\rm col}$ to the equilibrium temperature $T_{\rm equ}$ at the heliocentric distance $r_h$ of the comet:

\begin{equation}\label{eq:1}
S_{\rm heat} = \frac{T_{\rm col}}{T_{\rm equ}} = \frac{T_{\rm col}}{278 r_h^{-0.5}}.
\end{equation}

The bolometric albedo is defined as :

\begin{equation}\label{eq:2}
A(\theta) = \frac{f(\theta)}{1+f(\theta)}
\end{equation}

\noindent
with

\begin{equation}\label{eq:2b}
f(\theta) = \frac{[\lambda F_{\rm scatt}(\lambda)]^{\lambda_{\rm
max, scatt}}}{[\lambda F_{\rm therm}(\lambda)]^{\lambda_{\rm max,
therm}}}.
\end{equation}

The quantities $[\lambda F_{\rm scatt}(\lambda)]^{\lambda_{\rm max}}$ and $[\lambda F_{\rm therm}(\lambda)]^{\lambda_{\rm max}}$ are the apparent intensities (times the wavelength) measured at the maxima of the Planck function in the scattered and thermal part of the spectrum, respectively. $A$($\theta$) is referred to as the bolometric albedo, since it is approximately equal to the ratio between the scattered energy by the coma to the total incident energy \citep{Woodward2015}.

The parameters $S_{\rm heat}$ and $A$($\theta$) have been measured in several comets \citep[e.g.,][]{Mason2001},  including during outbursts \citep{Yang2009}, and can be retrieved from VIRTIS-H spectra (Fig.~\ref{fig:spectra}) by fitting the SED by the sum of the solar radiation scattered by coma dust grains, and a variable temperature blackbody describing the thermal emission from absorbing grains.

We modelled the scattered flux as:

\begin{equation}
F_{\rm scatt}(\lambda) = \frac{F_{\odot}(\lambda) g_{\rm
col}(\lambda) K_{\rm scatt}}{r_{\rm h}^2},
\end{equation}

\noindent where $F_{\odot}(\lambda)$ is the solar irradiance at
$r_{\rm h}$ = 1 AU.  $K_{\rm scatt}$ is a coefficient determined
in the fitting process which is function of both the total
geometric cross-section of the dust grains within the FOV and the
scattering properties of the individual grains, such as the
scattering phase function, and their scattering efficiency (Appendix
\ref{app:emission}). $g_{\rm col}(\lambda)$ is a factor which
describes the wavelength dependence of dust scattering. $g_{\rm
col}(\lambda)$ is expressed as:


\begin{equation}
g_{\rm col}(\lambda) = (\frac{\lambda}{\lambda_{\rm
ref}})^{-\alpha_c}
\end{equation}


\noindent where $\lambda_{\rm ref}$ is taken equal to 2 $\mu$m.
From the spectral index $\alpha_c$, one can deduce the
reflectivity gradient $S'_{\rm col}$, also called the dust
colour and usually expressed in \% per 100 nm. VIRTIS-H spectra
provide information on the spectral index between 2--2.5 $\mu$m.
The dust colour in this range is computed according to:

\begin{equation}\label{eq:5}
S'_{\rm col} = (2/500) \times \frac{g_{\rm col}(2.5\mu m)-g_{\rm
col}(2.0\mu m)}{g_{\rm col}(2.5\mu m)+g_{\rm col}(2.0\mu m)}
\end{equation}

The thermal radiation is modelled as:

\begin{equation}\label{eq:6}
F_{\rm therm}(\lambda) = B(\lambda,T_{\rm col}) K_{\rm therm},
\end{equation}

\noindent
where $B(\lambda,T_{\rm col})$  is the Planck function, and  $T_{\rm col}$ and $K_{\rm therm}$ are free parameters determined by the fitting routine.  $K_{\rm therm}$ is the absorption cross-section which depends  on the absorption properties of the individual grains, and the emitting cross-section within the FOV. In calculations this parameter is assumed to be constant in the wavelength range covered by VIRTIS-H.


The fit of the observed spectra was performed by means of the $\chi^2$ minimisation algorithm of Levenberg-Marquardt, with altogether up to five free parameters: the dust parameters $K_{\rm scatt}$, $K_{\rm therm}$, $T_{\rm col}$, and $\alpha_c$, and a multiplying coefficient $K_{\rm H_2O}$ to a synthetic water emission profile computed as described in \citet{dbm2016} which was added to the "scattering+thermal" dust model. Indeed, the water emission band at 2.7 $\mu$m occupies a large fraction of the spectrum in the range where both the scattered and thermal dust emissions are significant (Fig.~\ref{fig:spectra}). The spectral regions covering the CO$_2$ and $^{13}$CO$_2$ bands  (4.2--4.5 $\mu$m) and 
the CH stretches (3.3--3.6 $\mu$m) were masked in the fitting process.



The wavelength range of VIRTIS-H does not cover the maxima of the scattered and thermal dust spectra. Therefore, our determination of $A$($\theta$) is done using extrapolation. For computing $[F_{\rm scatt}(\lambda)]^{\lambda_{\rm max}}$ ($\lambda_{\rm max} \sim $ 0.5 $\mu$m), we took into account the change of the dust colour with wavelength, and assumed $S'_{\rm col}$ = 11\%/100 nm from 0.5 to 0.9 $\mu$m and = 1.7 \%/100 nm from 0.9 to 2.0 $\mu$m, as determined by \citet{Rinaldi2016} for comet 67P at $r_{\rm h} \sim$ 1.8--2.2 AU. The maximum of the dust thermal emission was derived from the maximum of $F_{\rm therm}(\lambda)$ (Eq.~\ref{eq:6}), using the derived $T_{\rm col}$ and $K_{\rm therm}$. This approach provides approximate values for $A$($\theta$). Concerning the thermal part, the colour temperature is a function of wavelength as a result of wavelength-dependent and size-dependent dust emissivities, whereas we are assuming that $T_{\rm col}$ is constant. This might affect retrievals when $T_{\rm col}$ is low, the Planck function peaking then well above 5 $\mu$m ($\sim$ 10 $\mu$m for 300 K). Concerning the reflected part, sources of errors are the effective dust colours from 0.5 to 2 $\mu$m, which may differ from the assumptions, especially for spectra acquired during the outbursts. From VIRTIS-M data, the dust colour measured between 0.45--0.75 $\mu$m remained in the range 10--15 \%/100 nm  and 6--12 \%/100 nm  during the time interval corresponding to the 13 and 14 Sept. VIRTIS-H data, respectively \citep{Rinaldi2017}, with the lowest values corresponding to the outburst material. However, no data are available concerning the dust colour from 1 to 2 $\mu$m. Therefore, checks were done by
comparing the extrapolated $F_{\rm scatt}$(586 nm) to measurements obtained at 586 nm using VIRTIS-M at the same location and time than VIRTIS-H. We concluded that the bolometric albedos derived by our method are correctly estimated.



\subsection{Data analysis}
\label{sec:reduction}
\begin{figure}
    \includegraphics[width=\columnwidth]{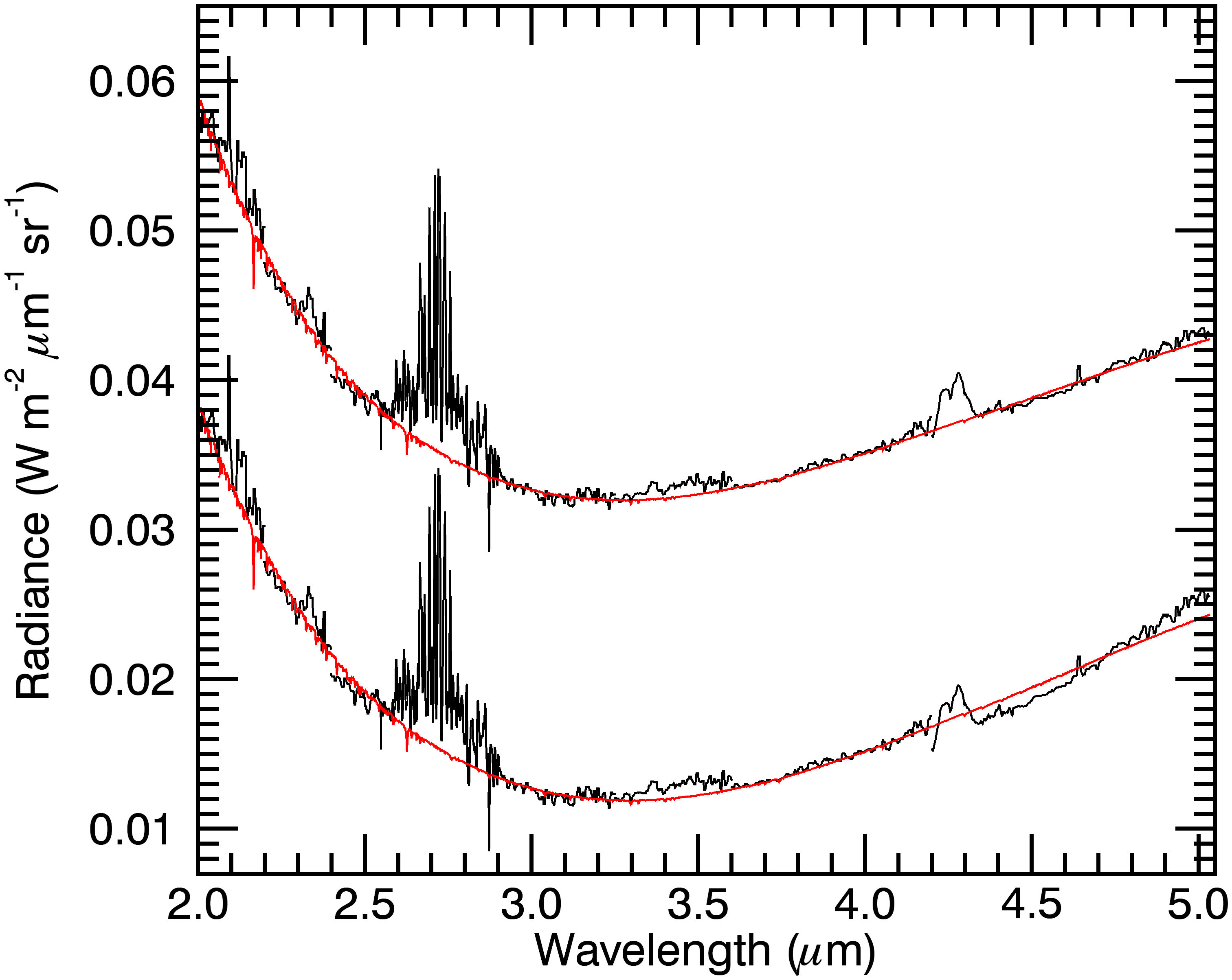}
       \caption{VIRTIS-H spectrum acquired on 13 September from 13.7125 to 13.7968 h UT. In the top spectrum, radiance values measured in the range 4.2--5 $\mu$m were corrected as described in Sect.~\ref{sec:reduction}, with $s$=0.2. This correction is not applied in the bottom spectrum. Overplotted, in red, are the best model fits.      }
    \label{fig:spectra2}
\end{figure}

Model fitting was applied to spectra obtained by averaging several acquisitions (6 to 11 at times showing outburst-related excess signal, up to 50 during quiescent states).
VIRTIS-H spectra are sampled over 3456 spectral channels, 432 in each grating order. Since adjacent orders cover common wavelength ranges, we have carefully selected the most reliable channels. In particular,
the short wavelength range of each order is often contaminated by instrumental straylight \citep{dbm2016} and was not considered.


Resulting spectra present inconsistent radiance values in the
overlapping regions of the diffraction orders. The discrepancies
are within 15\% for non-outburst data, and within 10\% for
outburst data, and cannot be only attributed to instrumental
straylight. For example, the mean intensity between 2.85--2.95
$\mu$m is lower for order 2 than for order 3,  whereas straylight
would produce the opposite. More problematic is the junction
between orders 0 and 1 (4.2 $\mu$m). For acquisitions acquired
during the outbursts (i.e., higher fluxes), radiance values are
lower in order 0 than in order 1, and the discrepancy between the
two orders increases with increasing signal (reaching values up to
10\%). In addition, model fitting shows that the continuum
emission above 4.2 $\mu$m (i.e. in order 0) is distorted, as
illustrated in Fig.~\ref{fig:spectra2} for a spectrum acquired
during an outburst. The distortion is akin to a change of the
bending of the spectrum, resulting in a higher flux than expected
above $\sim$ 4.5 $\mu$m, and a lower flux than expected below 4.6
$\mu$m. This effect is weaker, but still present in non-outburst
data. The origin of this likely instrumental distortion is
unclear. To correct the spectra, radiance values above 4.2 $\mu$m
were divided by a factor equal to (1--$\lambda_{ref}\times s$)+$s
\times \lambda$, $\lambda_{ref}$ = 4.5 $\mu$m. For $s$ = 0.2, this
factor increases the radiance at 4.2 $\mu$m by a factor of 1.06,
decreases the radiance at 5 $\mu$m by a factor of 1.10, whereas
radiance values near 4.5 $\mu$m remain unchanged. Data acquired
outside the outbursts suggest values $s$=0.1--0.2, whereas for
outburst spectra $s$=0.2--0.4. Model fitting to spectra with this
correction applied leads to higher values for $T_{\rm col}$ and
$A$($\theta$), and lower values for $S'_{\rm col}$.
Uncertainties given in Table~\ref{tab:properties} take into
account the range of values obtained for $s$=0 to 0.2 (appropriate
for 14 Sept. data and pre-outburst data), and $s$=0.2 to 0.4 (13
outburst Sept. data). Other results presented in the paper come
from spectra corrected with $s$=0.2.


\begin{figure}
   \centering
    \includegraphics[width=7.9cm]{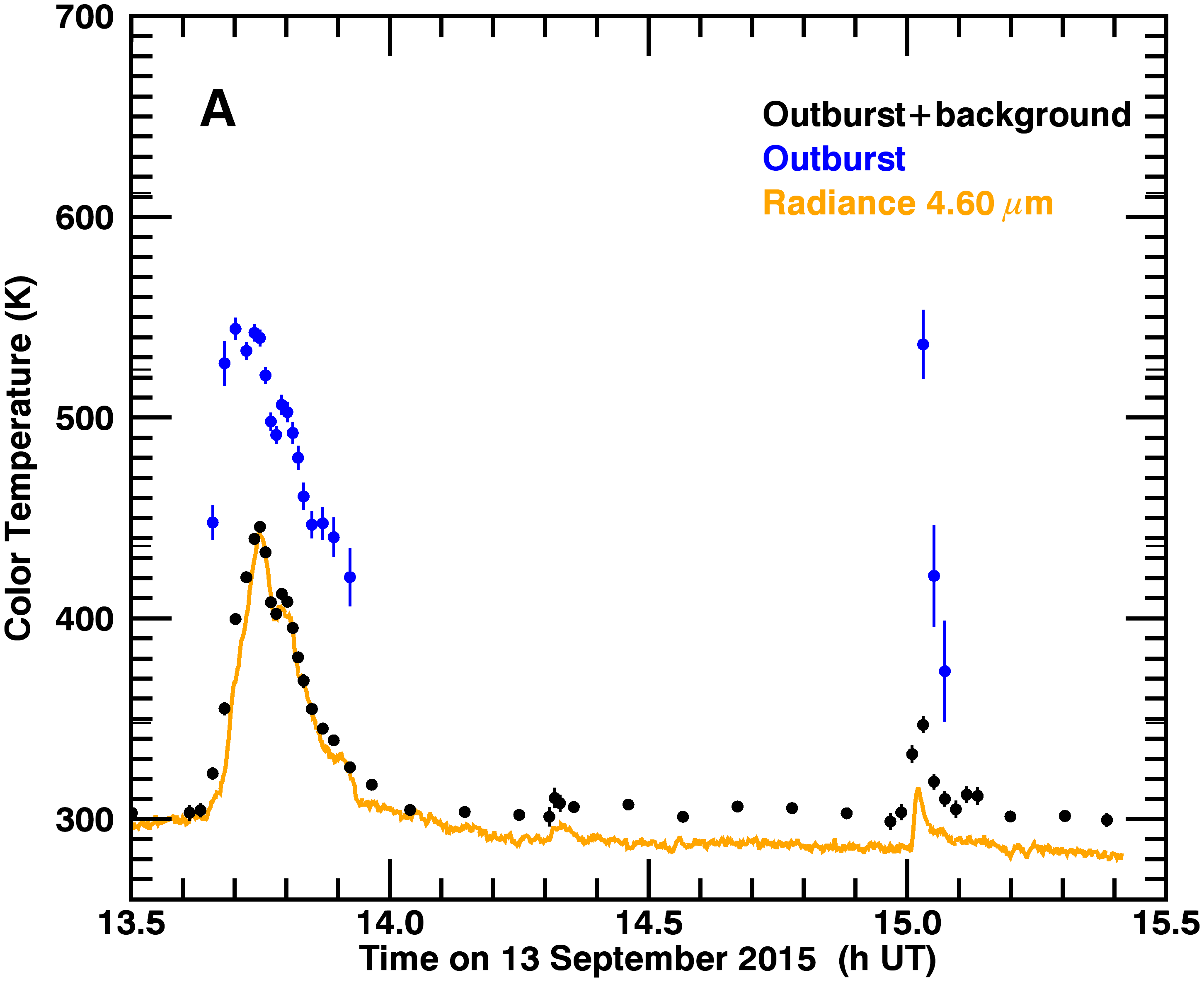}
    \includegraphics[width=7.9cm]{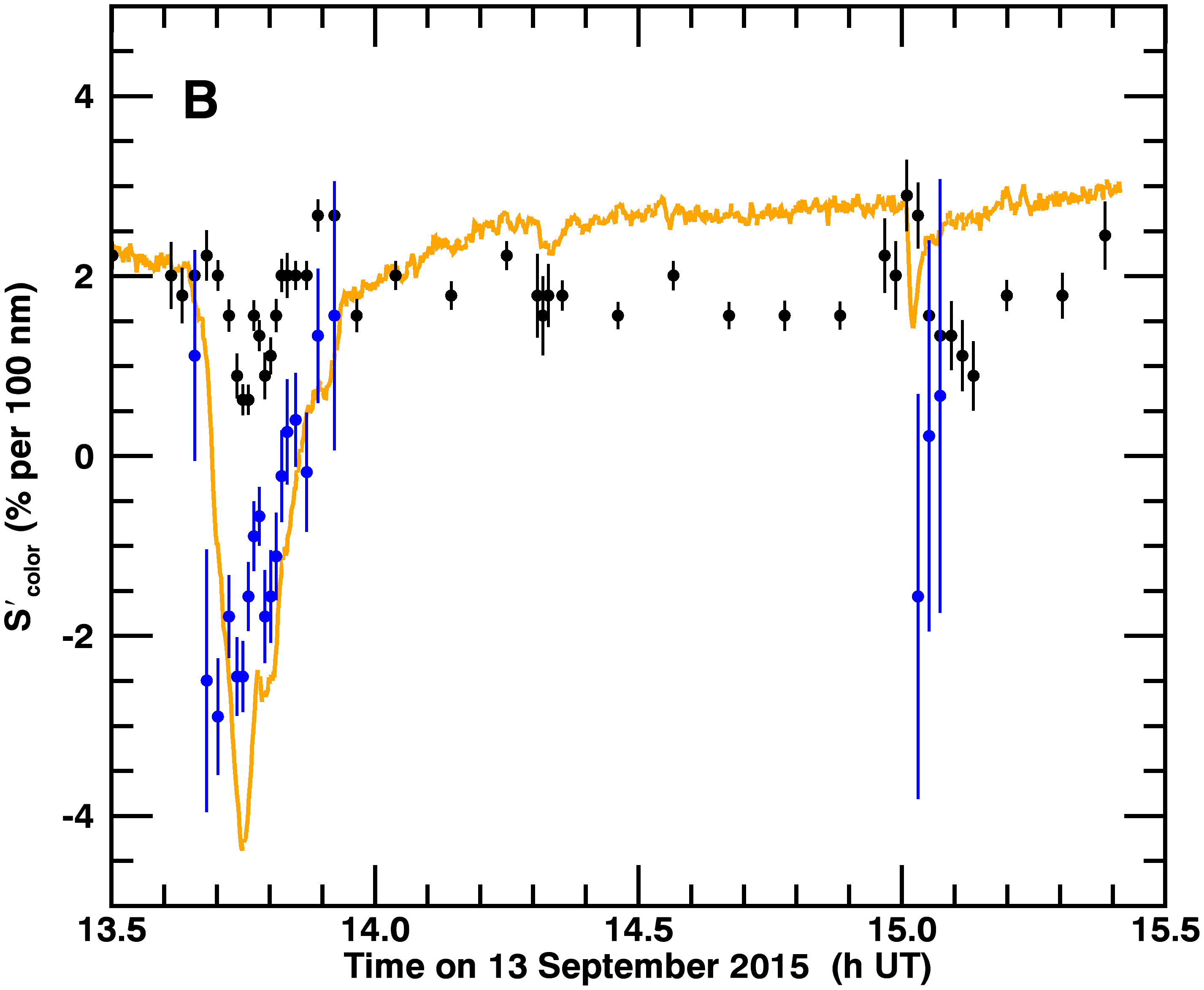}
    \includegraphics[width=7.9cm]{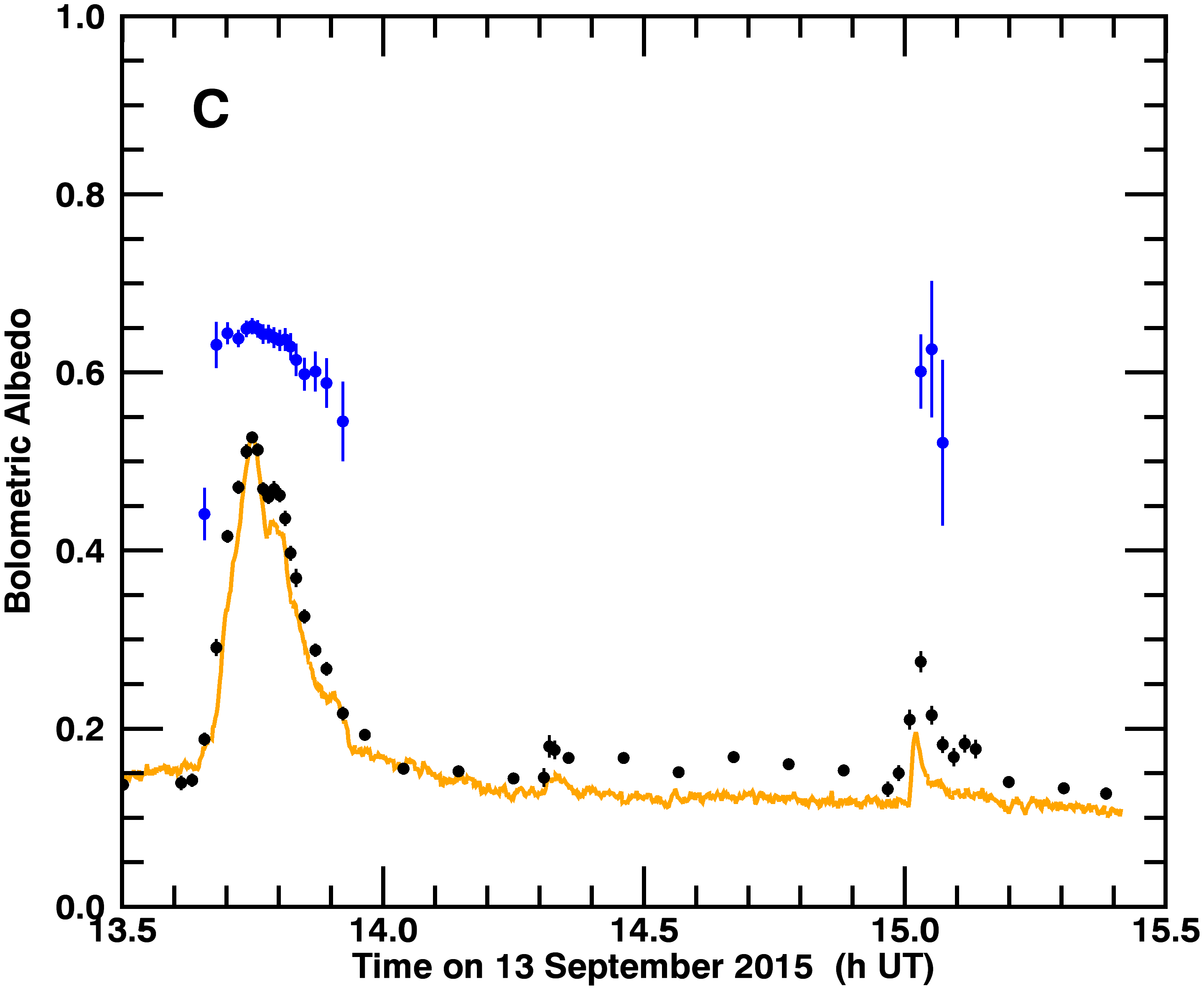}
\caption{Colour temperature (A), 2--2.5 $\mu$m colour (B) and bolometric albedo (C) on 13 September 2015. Black symbols correspond to retrievals from total spectra, whereas blue symbols are for spectra with a reference pre-outburst spectrum removed. The radiance measured at 4.6 $\mu$m (with adapted scale) is shown in orange colour.    }
    \label{fig:properties-13SEPT}
\end{figure}


Some despiking was applied by ignoring spectral channels
with a residual intensity with respect  to a rough model fit
larger than two times the rms noise. The rms noise corresponds to
the standard deviation of radiance values measured across the
residual spectrum. Since the statistical noise varies along the
spectrum (data at short wavelengths are more noisy, see
Fig.~\ref{fig:spectra}), it was computed over spectral windows,
and assumed to be representative of the noise in these windows.
The model has five
free parameters. To find the optimum parameters, we computed
contours for the variation of $\chi^2$ as a function of $T_{\rm
col}$ and $\alpha$, using fine gridding. For each pair of values
($T_{\rm col}$, $\alpha$), the optimum values of $K_{\rm
scatt}$, $K_{\rm therm}$ and $K_{\rm H_2O}$ were determined using
the $\chi^2$ minimisation algorithm of Levenberg-Marquardt. The
best model fit is obtained for parameters corresponding to the
minimum $\chi^2$ in the ($T_{\rm col}$, $\alpha$) parameter
space. The 1--$\sigma$ confidence level for these two parameters
is derived from the contour at $\Delta\chi^2$ = 2.3, as described
in \citet{Bevington}. Uncertainties in $A$($\theta$) were computed
taking into account the uncertainties in $K_{\rm therm}$, $K_{\rm
scatt}$, and $T_{\rm col}$.

In order to best characterise the properties of the outburst
material, a reference spectrum acquired before the beginning of the outburst was subtracted to the spectra.
The first mini outburst of 13 Sept. was not studied in detail,
because of its faintness.

\begin{figure}
    \includegraphics[width=7.9cm]{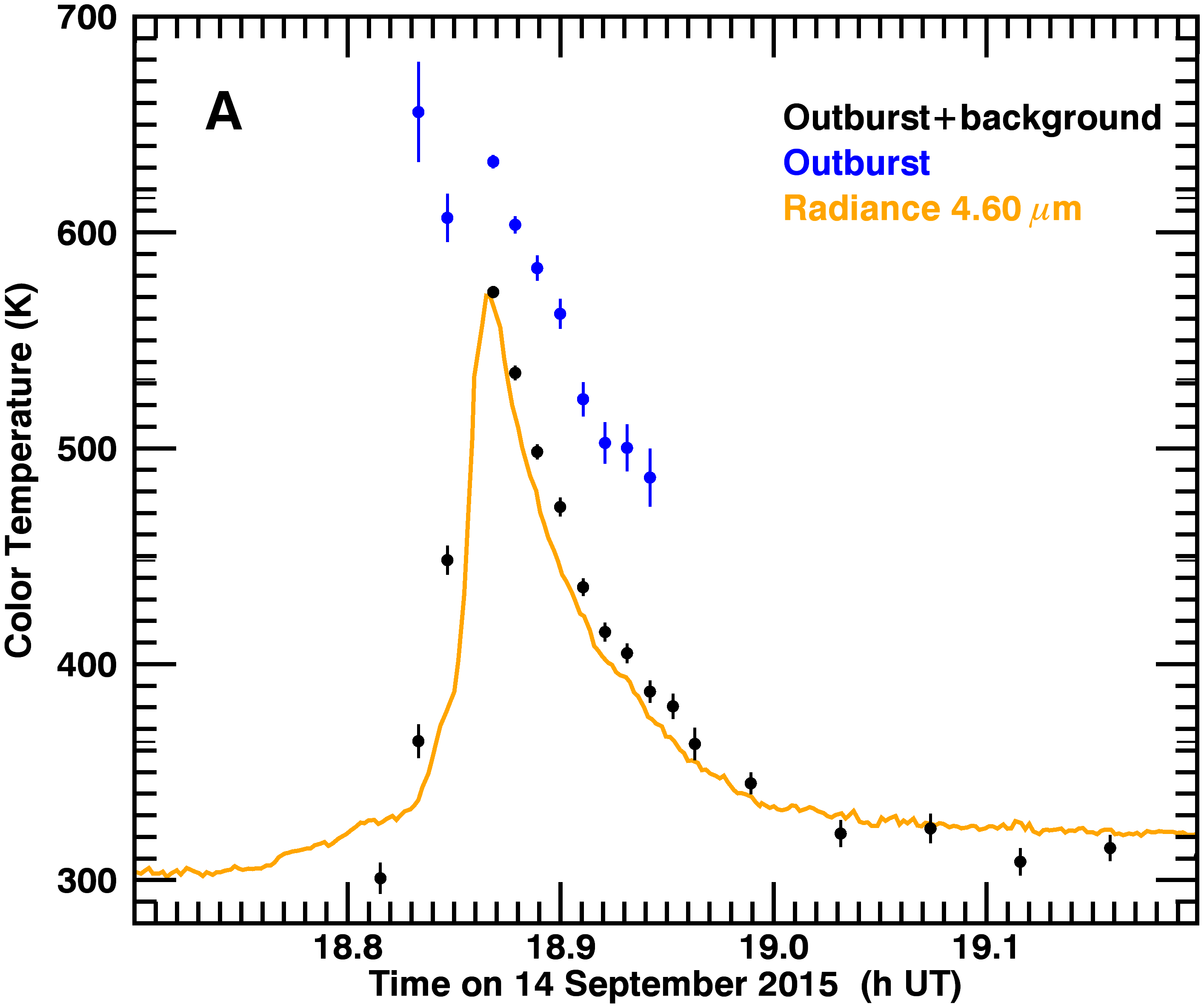}
    \includegraphics[width=7.9cm]{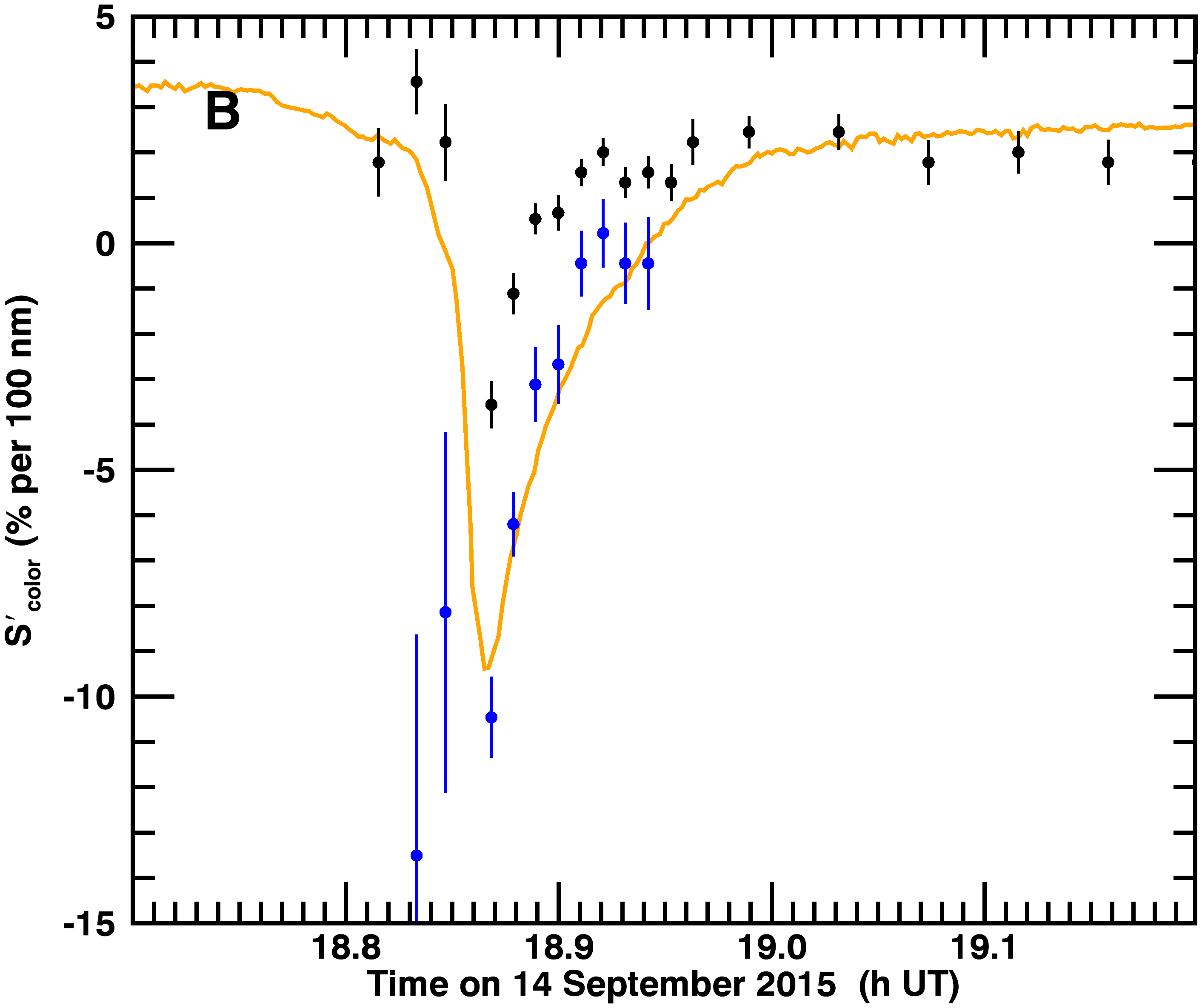}
    \includegraphics[width=7.9cm]{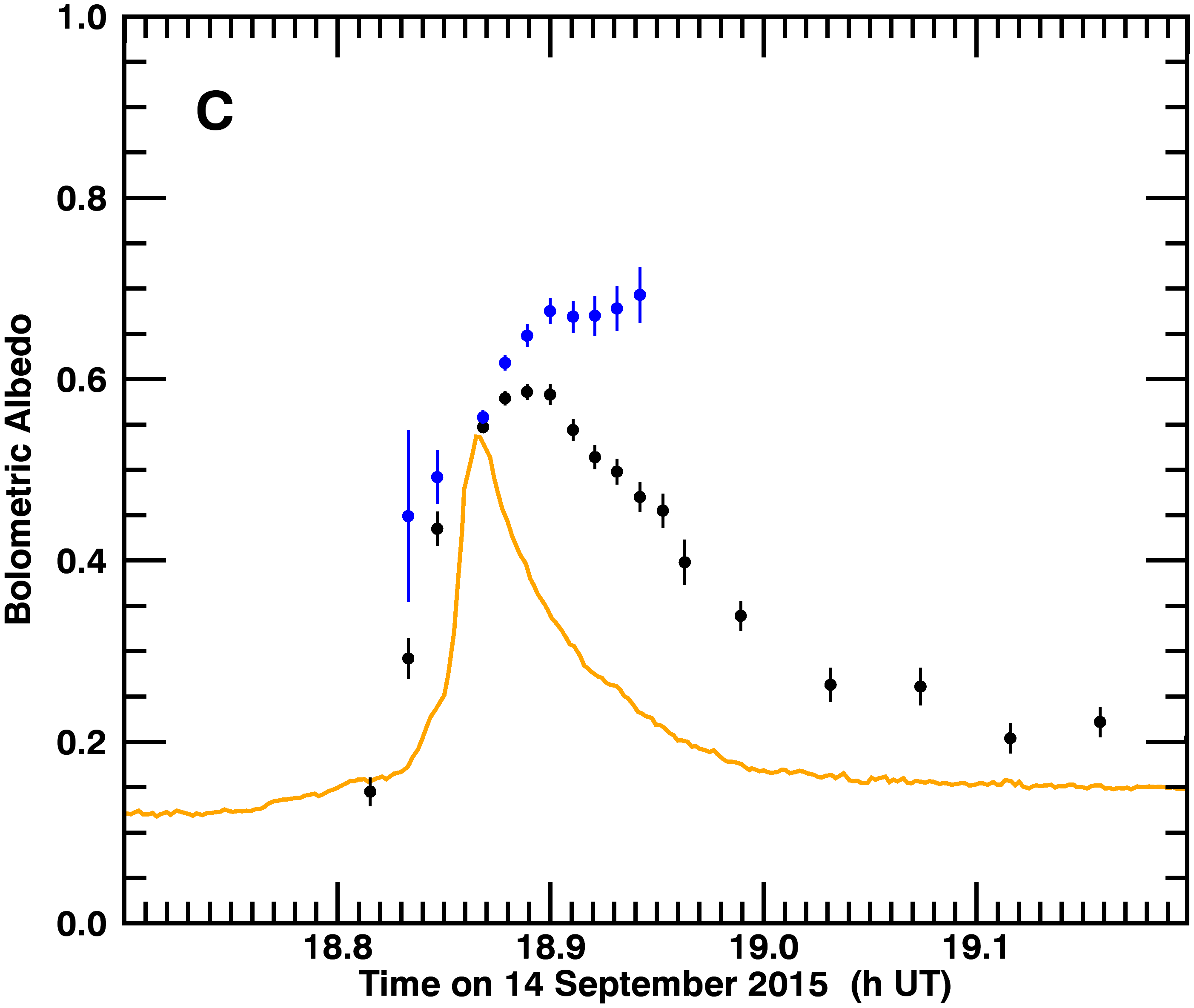}
\caption{Same as Fig.~\ref{fig:properties-13SEPT} at the time of the outburst of 14 September 2015.}
    \label{fig:properties-14SEPT}
\end{figure}

Figures~\ref{fig:properties-13SEPT} and ~\ref{fig:properties-14SEPT} show the
temporal evolution of $T_{\rm col}$, $S'_{\rm col}$ and
$A$($\theta$) retrieved from model fitting, with selected values reported in Table~\ref{tab:properties}.

\subsection{Retrieved dust scattering and thermal properties}
\label{sec:properties}

\subsection{Quiescent coma}
\label{sec:properties-back}

Spectra obtained just before the outbursts yield $T_{\rm col}$ $\sim$ 300 K, $A$($\theta$) = 0.13 and $S'_{\rm col}$ $\sim$ 2.5 \%/100 nm (Table~\ref{tab:properties}), and similar values are measured when the outburst material has left the FOV.

\begin{figure}
    \includegraphics[width=8.cm]{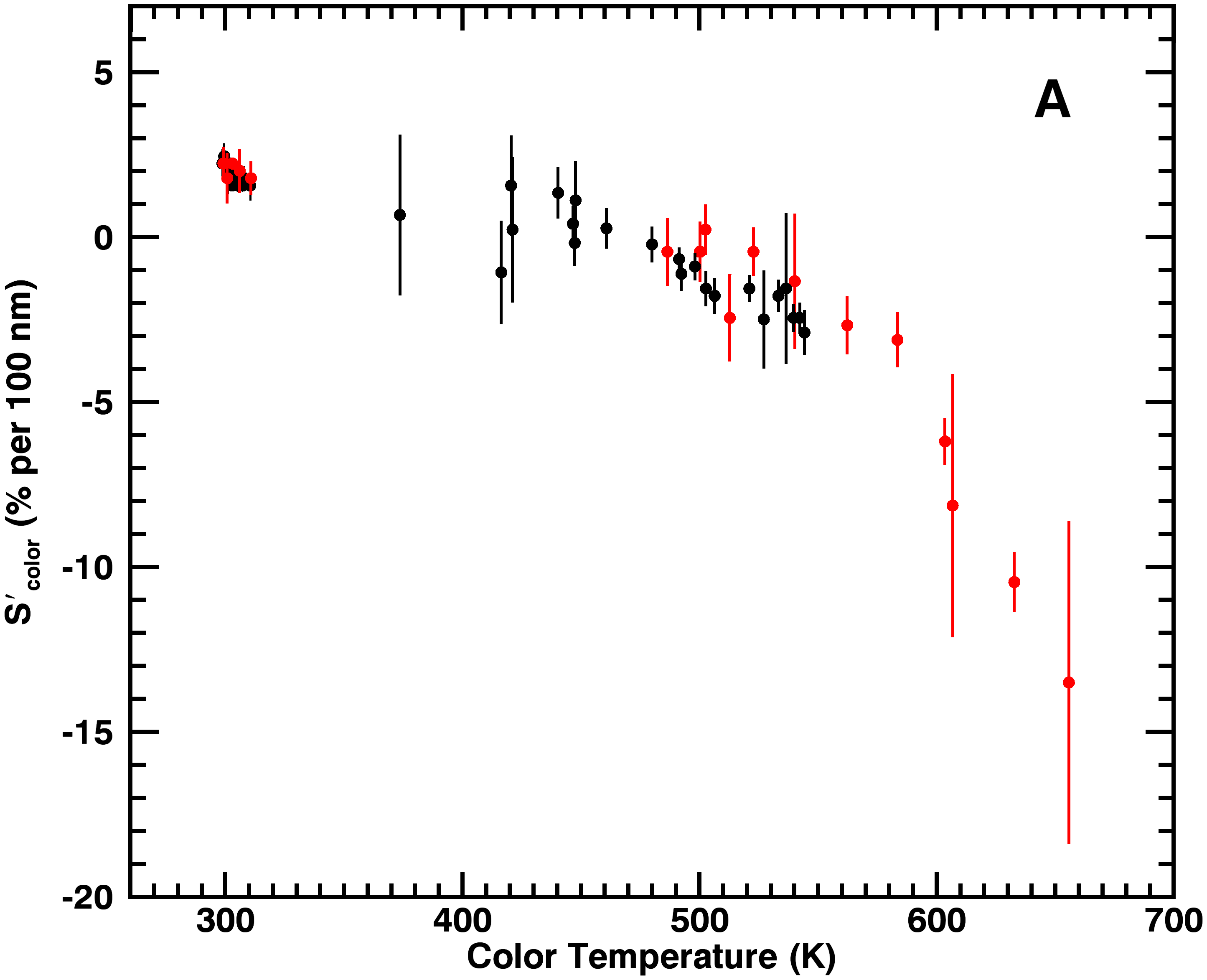}
       \par
\vspace{0.3cm}
    \includegraphics[width=8.cm]{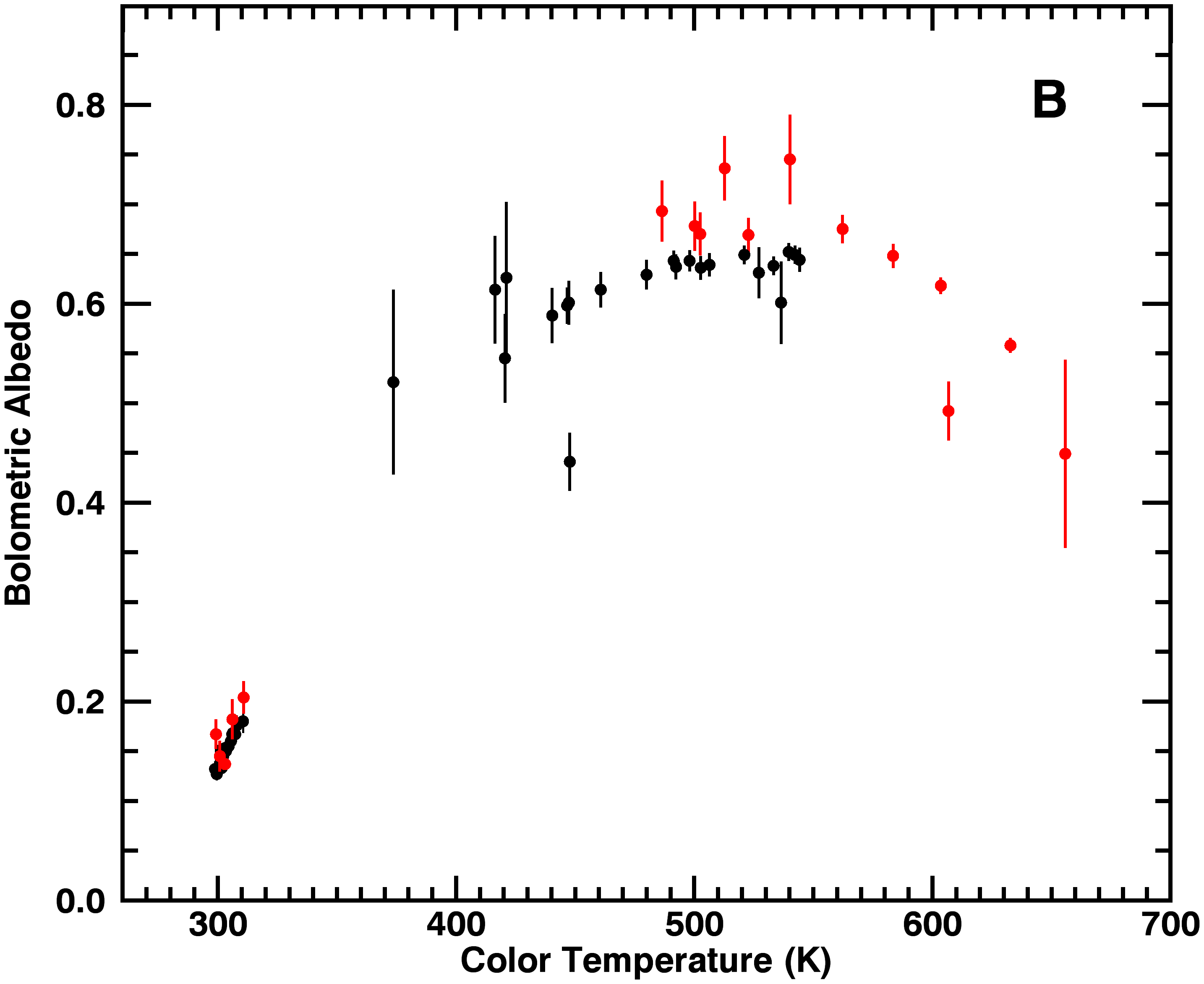}
\par
\vspace{0.3cm}
    \includegraphics[width=8.cm]{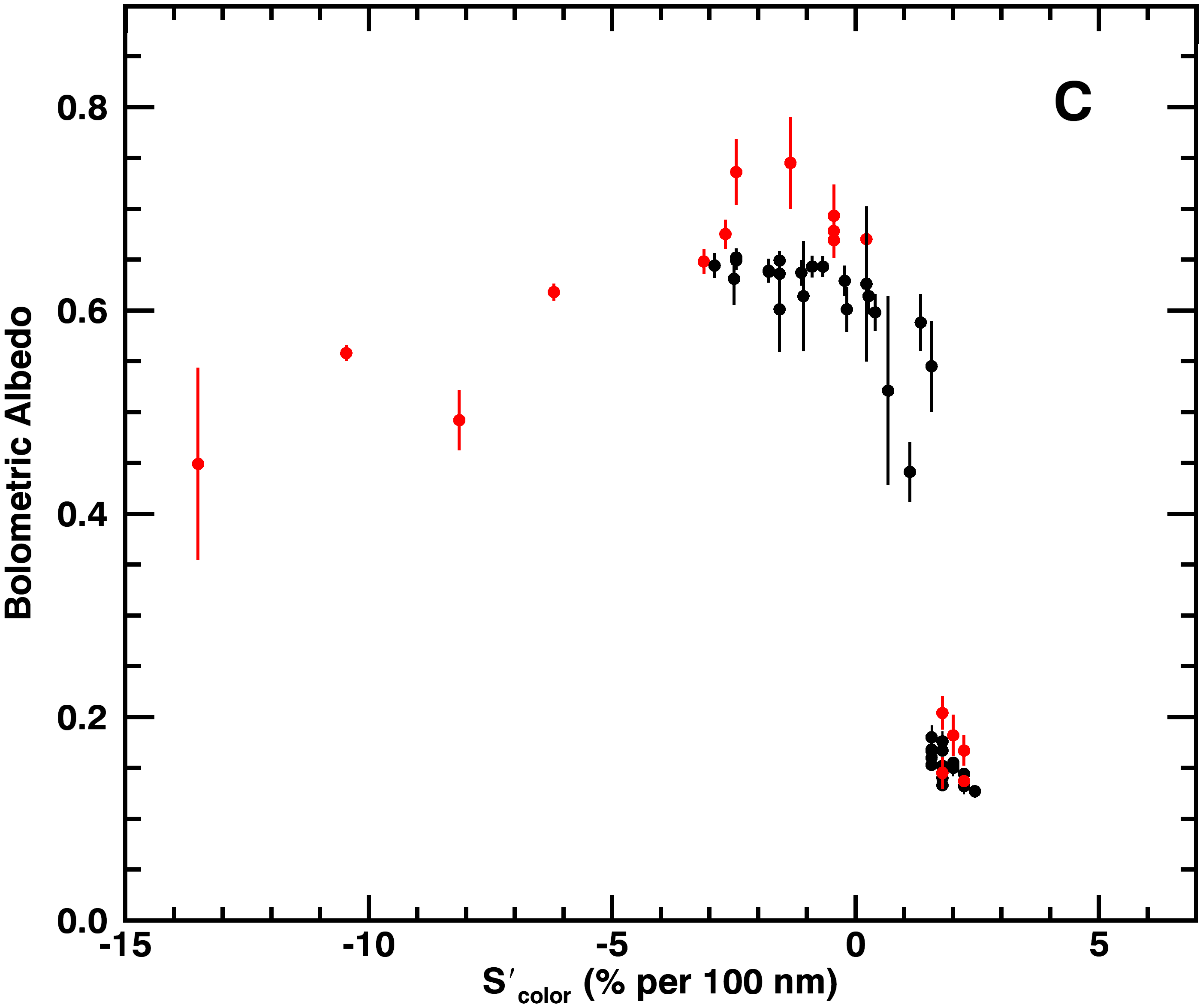}
     \caption{Correlation trends between colour temperature ($T_{\rm col}$), 2--2.5 $\mu$m colour ($S'_{\rm col}$), and bolometric albedo ($A$($\theta$)). Black and red symbols are for 13 Sept. and 14 Sept. data, respectively. }
    \label{fig:correlation}
\end{figure}

A quantity used to characterise the scattering of cometary dust, and measured in numerous comets \citep{Kolokolova2004} is the
geometric albedo $A_{\rm_p}$ times the phase function at phase angle $\alpha$ : $A_{\rm_p}$($\alpha$) = $A_{\rm_p}$ $\times$ $j$($\alpha$), with $j$(0) = 1 (see Sect. \ref{sec:properties}). The total dust cross-section is derived from the thermal SED assuming an emissivity equal to 1, and this cross-section is then applied to the scattered intensity to derive an average $A_{\rm_p}$($\alpha$). We used this approach
to derive $A_{\rm_p}$($\alpha$) at $\lambda$ = 2 $\mu$m from the VIRTIS-H spectra.
Retrieved values are $A_{\rm_p}$($\alpha$) = 0.018 and 0.021 for the pre-outburst spectra of 13 and 14 Sept., respectively. A plot of $A_{\rm_p}$($\alpha$) as a function of time, with the same time sampling as in Figs~\ref{fig:properties-13SEPT}--\ref{fig:properties-14SEPT} shows temporal variations.  Considering only data points  outside periods showing outburst-related signal excess, we find an average value $A_{\rm_p}$($\theta$)= 0.020 $\pm$ 0.004, where the error bar corresponds to the standard deviation of the sample.
For these data points, we observe a positive correlation between $A_{\rm_p}$($\alpha$) and $T_{\rm col}$, with values of 0.025 reached for $T_{\rm col}$ = 308--310 K. A positive correlation with $T_{\rm col}$ is also observed for the bolometric albedo
(Fig.~\ref{fig:correlation}B), from which we conclude that some data points  are affected by outburst-related material.

\subsection{Outbursts}
The scattering and thermal properties of the dust particles in the
outburst material strongly differ from those of the background
coma (also referred to as the quiescent coma in this paper).  The first acquisitions taken after the onset of the
outbursts show a large increase in the colour temperature and bolometric albedo, and less red or bluish dust colours. Spectra of the outburst material (i.e., with
the background reference spectrum removed) show colour temperatures reaching $\sim$ 550 K for the
two investigated outbursts of 13 Sept. (main and second mini ones), and $\sim$ 640 K for the 14. Sept. outburst. The three investigated outbursts all show negative blue colours, with an extreme value
$S'_{\rm col}$ $\sim$ --10 \%/100 nm measured at the peak of the 14 Sept. outburst corresponding to a power index for the 2--2.5 $\mu$m reflectance of $\alpha_{\rm c}$ = 2.30. They also show similarly high bolometric albedos
$A$($\theta$) $\sim$ 0.6. The three parameters show different behaviors as a function of time. The evolution of $T_{\rm col}$ and $S'_{\rm col}$ follow that of the light curves. They reach their most extreme values approximately at the time of the peak values of the radiances, returning then progressively to pre-outburst values. Several (at least three) local extrema are observed for these parameters during the 13 Sept. main outburst (Fig.~\ref{fig:properties-13SEPT}), confirming that the complex light curve of this event results from consecutive outbursts.
Regarding $A$($\theta$), its temporal evolution follows that of light curves for the 13 Sept. outbursts only. For the 14. Sept. outburst material, measured $A$($\theta$) values remain high (and are even still increasing) after the decay of brightness has proceeded. This trend is at the origin of the very distinct light curves in the thermal and scattered wavelength ranges (Fig.~\ref{fig:lightcurve}).

To what extent are the light curves influenced by optical depth effects in the coma? 
The opacity can be estimated by ratioing measured radiances to the value expected for a fully opaque cloud. Assuming that the product of the normal geometric albedo times the phase function is equal to 0.02, as measured for the quiescent coma 
(Sect.\ref{sec:properties-back}), we find $\tau$ = 0.1 at 2$\mu$m
for the peak brightness on 14 Sept.. This is a lower limit to the
opacity, since outburst ejecta consist of bright dust particles (Sect.~\ref{sec:outburst}). The fact that light curves do not stall at their maximum value
during some time interval is  suggesting that opacity effects are
not significant.

The temporal variability of the spectral properties is related to changes in the properties of the dust outburst material in the course of the event. The mean speed of dust grains detected by the Grain Impact Analyser and Dust Accumulator (GIADA) during the 19 Feb. 2016 outburst is 6.5 m s$^{-1}$, pertaining to a mean radius of 180 $\mu$m \citep{Gruen2016}. If we assume a speed dependence in $a^{-0.5}$, then grains with radii smaller than 10-$\mu$m
(those contributing mainly to IR emission) have speeds exceeding 30 m s$^{-1}$. Their travel time in the VIRTIS-H FOV is $\leq$ 6 s. Therefore, consecutive acquisitions do not sample the same material (providing  projection effects are minimal). In Figs~\ref{fig:properties-13SEPT}--\ref{fig:properties-14SEPT}, the temporal evolution of the dust properties is presented with a time resolution $\geq$ 40 s, therefore each data point should pertain to distinct material.

\section{Interpretation with light-scattering modelling}
\label{sec:analysis}

   \begin{figure}
    \includegraphics[width=8cm]{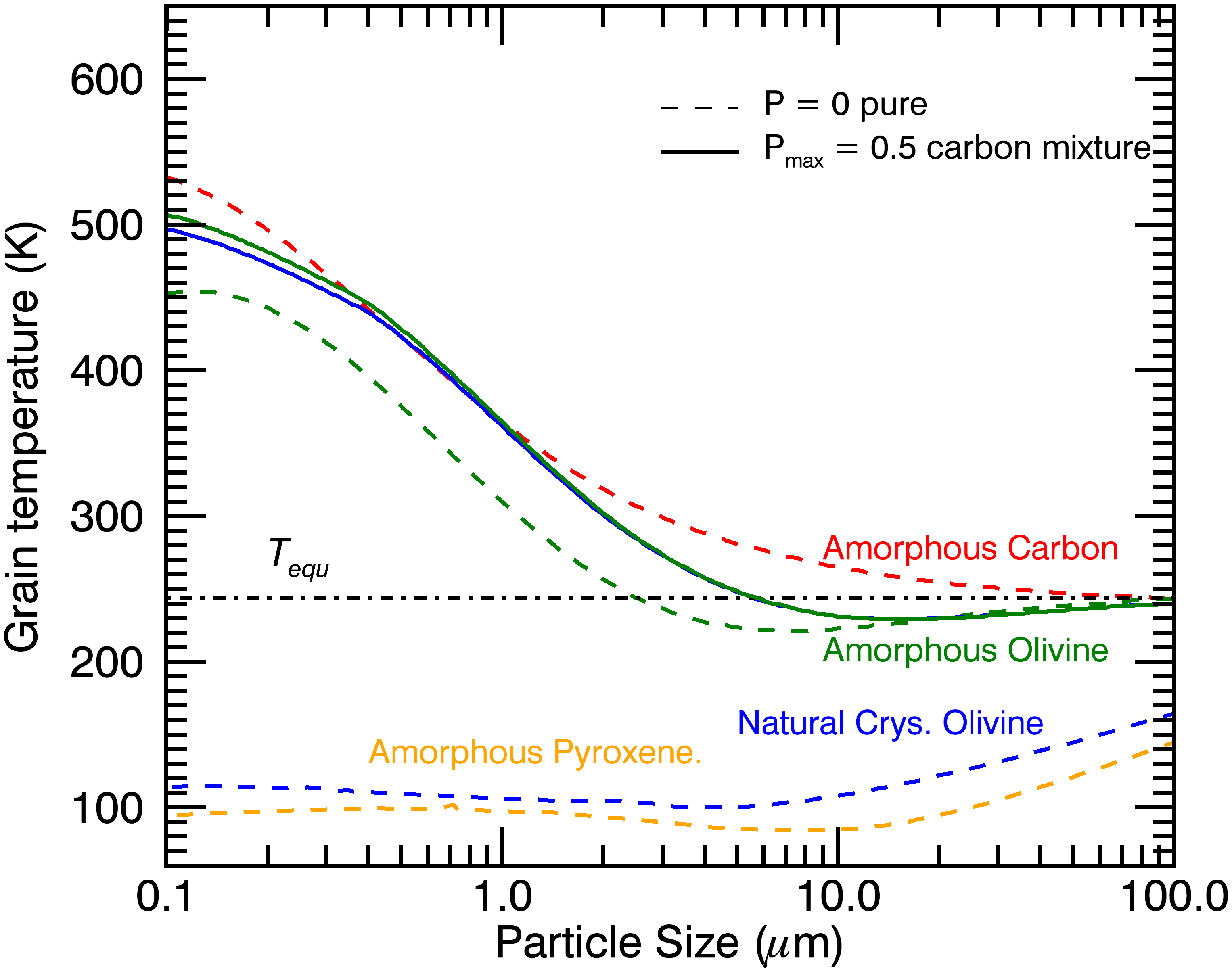}
\caption{Grain temperature as a function of grain radius at $r_{\rm h}$ = 1.3 AU.
Dashed curves are for non-porous grains ($P$ = 0) made of unique material: amorphous carbon (red),
natural Mg-rich crystalline silicates (blue), amorphous MgSiO$_3$ pyroxene (orange) and amorphous olivine with Mg:Fe = 50:50 (green). Plain curves are for moderately porous ($P_{\rm max}$ = 50\%, $a_{\rm unit}$ = 100 nm, $D$ = 2.5) carbon/silicate mixtures ($q_{\rm frac}$ = 0.7 corresponding to a silicate volume fraction of 34\%) with same colour coding. Temperatures for mixtures of carbon with Mg-rich crystalline silicates are the same as for mixtures of carbon with amorphous MgSiO$_3$ pyroxene, so the curves are superimposed. The equilibrium temperature $T_{\rm equ}$ is shown as a black dashed-dotted line.}
    \label{fig:Tdust}
\end{figure}

\subsection{Modelling}
\label{sec:model}

The scattering and thermal properties of cometary dust depend on the particle composition and structure, and on the size distribution \citep{Hanner2003,Kolokolova2004}. 
In order to analyse the 67P observations we used the Mie \citep{Mie1908} and Rayleigh-Gans-Debye (RGD) \citep[][and references therein]{Tazaki2016} scattering theories, combined with an effective medium theory in order to consider mixtures of different materials.

The GIADA experiment onboard the Rosetta spacecraft has suggested
that most of the dust released from 67P consists of relatively
compact particles with bulk densities between 10$^3$ and 3
$\times$ 10$^3$ kg m$^{-3}$  and a mean porosity $P$ = 0.5
\citep{Rotundi2015,Fulle2015,Fulle2016,Dellacorte2016}. To model
the infrared spectrum of these moderately porous grains, we used
the Mie scattering theory. Mie theory considers homogeneous
spheres, and provides reliable results in the Rayleigh regime
(i.e., for size parameter $x$ = 2$\pi a /\lambda$ $<<$ 1, where
$a$ is the particle radius), and for computing the emissivity and
temperature of dust grains \citep{Kolokolova2004}.

The GIADA instrument detected short-lasting showers of low-density
grains, resulting from the fragmentation of very fluffy,
fractal-like particles \citep{Fulle2015}. Fractal-like particles
were collected and imaged by the Rosetta's Micro-Imaging Dust
Analysis System (MIDAS) onboard Rosetta \citep{Mannel2016}. The
large fractal aggregate ($\sim$ 20 $\mu$m) examined by
\citet{Mannel2016} has a fractal dimension $D$ = 1.7 with subunits
of mean diameter $\sim$ 1.5 $\mu$m that are expected to contain
subunits of tenths of micrometre size \citep{Bentley2016}.
Modelling the scattering properties of aggregates can be achieved
by utilising numerical methods such as the T-matrix method (TMM,
for cluster spheres) or the discrete dipole approximation (DDA,
for arbitrary shape particles). These methods need time-consuming
computations, and because of this limitation, they have been
applied so far only to small aggregates of a few micrometre size
at most  \citep{Kolokolova2004,Kimura2016}. The analysis of the
VIRTIS spectra  would require considering large aggregates. We made simulations for fractal aggregates under the
hypothesis that the constituent subunits are Rayleigh scatterers
and that the aggregates have a Balistic Cluster Cluster Aggregate
(BCCA) topology. BCCA typically have a fractal dimension $D \leq$
2.0, and so resemble 67P fluffy grains. For such aggregates, the
Rayleigh-Gan-Debye (RGD) theory can be used \citep{Tazaki2016}. 
Appendix B provides details on the performed RGD calculations, done for  $D$
= 1.7 and a subunit size of 100 nm.

Effective medium theories (EMT) allow us to calculate an effective
refractive index for a medium made of a matrix (here assumed to be
amorphous carbon) with inclusions of different composition (e.g.,
silicates). We used the Maxwell-Garnett
mixing rule \citep{MG1904}. We defined the volume
fraction of the inclusions in dust grains by the quantity $q_{\rm
frac}^3$ (Appendix A).

In order to simulate the infrared continuum emission of cometary
dust, one has to compute the grain temperature as a function of
grain size, which requires  optical constants from the visible to
far-infrared ranges (Appendix A). We considered the following
materials for which these data are available: 1) amorphous carbon,
2) amorphous olivine, with a Fe:Mg composition of 50:50
consistent with the value in GEMS (Glass with Embedded Metal and
Sulfides) found in IDPs, and in ISM dust \citep{Keller2011}, 3)
amorphous Mg-rich pyroxene (Fe:Mg of 0:100), 4) Mg-rich
crystalline olivine (spectral indices from a natural olivine), and
5) troilite FeS. Amorphous carbon is a strongly absorbing material
in the visible and in the near-infrared. The refractory
carbonaceous matter in cometary grains is likely polyaromatic,
according to the known composition of chondritic porous
stratospheric IDPs, of fine-grained fluffy Concordia Antarctic
meteorites, and of STARDUST grains that were not harshly modified
during their capture in aerogel collectors \citep[][De Gregorio,
personal communication]{Sandford2006, Dobrica2011}. Polyaromatic
carbonaceous materials with small-sized polyaromatic units are
however poorly absorbent in the range 1-5 $\mu$m
\citep{Quirico2016}. However, carbonaceous matter in cometary
grains is mixed up with metal and pyrrhotite (Fe$_{\rm 1-x}$S),
which are believed to be at the origin of the low albedo of 67P
nucleus \citep{Capaccioni2015,Rousseau2017}. Here, we used the
optical constants of the less absorbent troilite \citep[][and
references therein]{Henning1996}, because no optical data are
available for pyrrhotite. The optical properties of amorphous
carbon and troilite finally fairly mimic a pyrrhotite+polyaromatic
carbonaceous matter mixture, which represents the opaque component
of cometary grains \citep[see][]{Engrand2016}. The silicate
component of cometary grains consists of crystalline and amorphous
olivines and pyroxenes
\citep{Wooden1999,Dobrica2011,Dobrica2012,Engrand2016}. The references for the
refractive indices are: amorphous carbon \citep{Edoh1983}, glassy
(amorphous) silicates \citep[namely MgSiO$_3$ pyroxene, and
MgFeSiO$_4$ olivine,][]{Dorschner1995}, Mg-rich natural
crystalline olivine \citep{Fabian2001}.

The nominal volume fraction of silicates in the grain is set to 0.34 ($q_{\rm frac}$ = 0.7), which is close to the value estimated for the grains of 67P by \citet{Fulle2016} (0.33 excluding ices).  The nominal model will consider amorphous olivine for the carbon/silicate mixture, which is relatively absorbent due to the presence of Fe, whereas other silicates are transparent.

In order to account for the porosity of the moderately compact grains, we followed the simple approach of \citet{Greenberg1990}, who applied the Maxwell-Garnett mixing rule with voids taken as the matrix and the mixture of constituent materials as the inclusions.  The porosity is described by a fractal dimension $D$ and the radius of the subunits $a_{\rm unit}$:

\begin{equation}\label{eq:poro}
P =1-(a/a_{\rm unit})^{D-3}.
\end{equation}

\noindent
We used the values $D$ = 2.5 and $a_{\rm unit}$ = 100 nm. This fractal dimension is consistent with the constraint obtained for 67P compact grains by MIDAS \citep{Mannel2016}. The maximum porosity of the grains is set to $P_{\max}$. Indeed, with the porosity law given by Eq.~\ref{eq:poro} the porosity increases with increasing size.  We set $P_{\max}$ = 0.5, corresponding to the value established for the so-called compact grains detected by GIADA \citep{Fulle2016}.


The size distribution of the dust particles is described by a power law  $n(a) \propto  a^{-\beta}$. $\beta$ is the size index and the particle radius $a$ takes values from $a_{\rm min}$ to $a_{\rm max}$. We set $a_{\rm max}$ = 500 $\mu$m (large particles do not contribute significantly to IR radiation). For sizes larger than 30 $\mu$m, the size index is set to $\beta$ = 1.9 \citep{Merouane2016}. We considered size distributions with $\beta$ values from 2 to 5 and $a_{\rm min}$ from 0.1 to 20 $\mu$m.

Grain temperatures and synthetic spectra were computed, as described in Appendices A--B.  The colour temperature, bolometric albedo and colour were retrieved from the synthetic spectra, as done for the VIRTIS-H spectra (Sect.~\ref{sec:approach}). The temperatures of compact ($P$ = 0) and moderately porous particles ($P_{\max}$ = 0.5) are plotted in Fig.~\ref{fig:Tdust}. Fractal aggregates with $D$ = 1.7 have temperatures equal to their constituent grains.
This well established trend \citep[][and references therein]{Kimura2016} is related to the fact that both the absorption cross-section and surface area of fractal agglomerates are proportional to the number of consituent grains in the limit of low fractal dimension. At 1.3 AU from the Sun, the temperature of fractal grains ($D$ = 1.7)
is $\sim$ 100~K for grains composed of transparent silicates and between 450--530 K for carbon-silicate mixtures and MgFeSiO$_4$ amorphous olivine (Fig.~\ref{fig:Tdust}).

\begin{figure}
   \includegraphics[width=8cm]{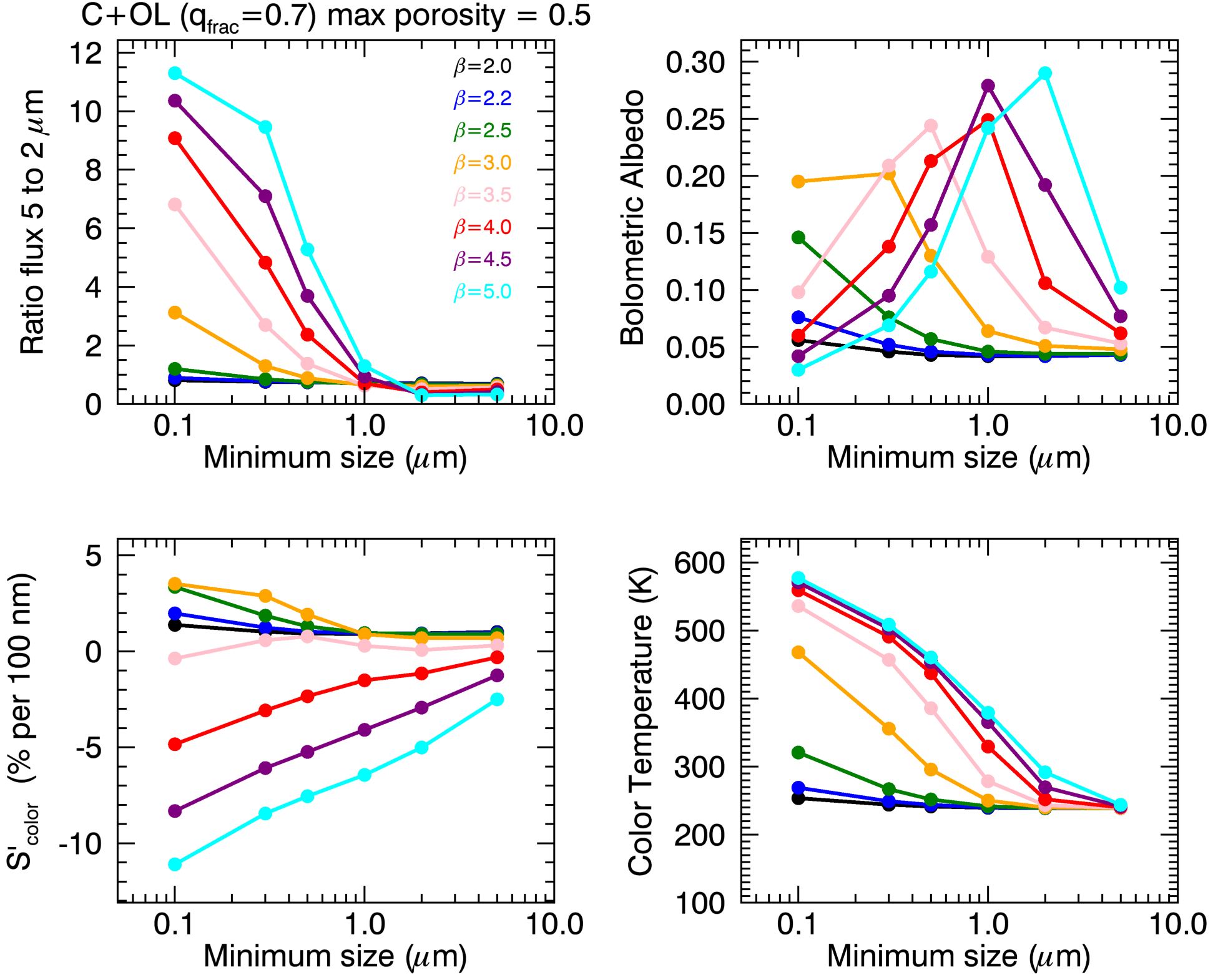}
    \caption{Results from Mie-scattering calculations for moderately porous ($P_{\max}$ = 0.5) carbon/olivine mixtures  as a function of minimum size $a_{\rm min}$. The different colours correspond to different size indices $\beta$ indicated in the top left figure. From left to right and top to bottom: ratio of the dust continuum radiation at 5 $\mu$m to the value at 2 $\mu$m, bolometric albedo, colour at 2 $\mu$m and colour temperature. The model parameters are: $q_{\rm frac}$ = 0.7, $a_{\rm unit}$ = 100 nm, $D$ = 2.5, phase angle $\alpha$ = 90 $^\circ$. The Mg:Fe content of olivine is 50:50.}
    \label{fig:Mie-Nominal}
\end{figure}

\begin{figure}
   \includegraphics[width=8cm]{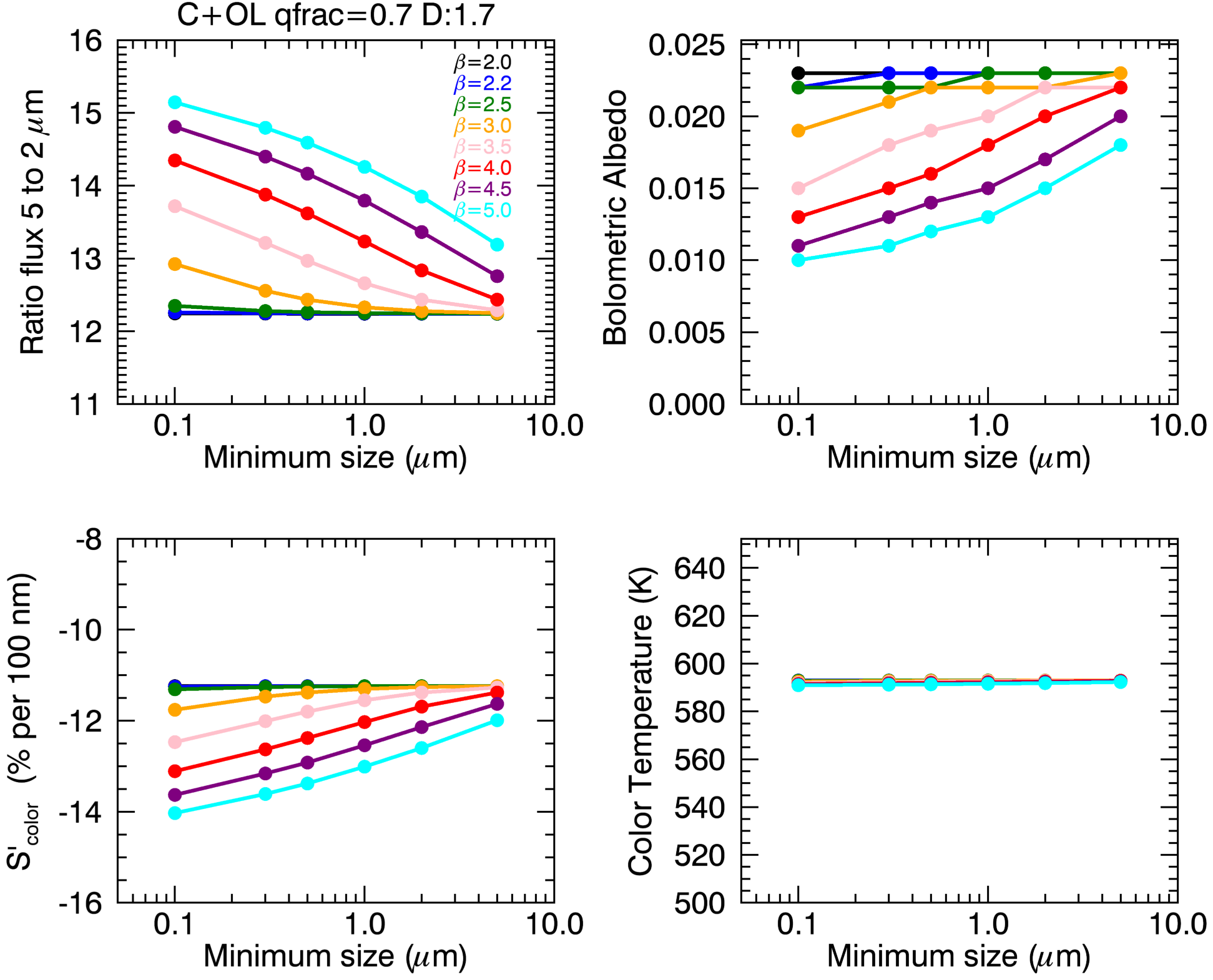}
    \caption{Results from Rayleigh-Mie-Debye calculations for fractal aggregates made of  carbon/olivine mixtures. See caption to Fig.~\ref {fig:Mie-Nominal}.
     The model parameters are: $q_{\rm frac}$ = 0.7, $a_{\rm unit}$ = 100 nm, $D$ = 1.7. The Mg:Fe content of olivine is 50:50.}
    \label{fig:RGD-Nominal}
\end{figure}

Spectral properties of the dust continuum radiation are plotted in Fig.~\ref{fig:Mie-Nominal}, for the moderately porous ($P_{\max}$ = 0.5) carbon-silicate mixtures with $q_{\rm frac}$ = 0.7, and in Fig.~\ref{fig:RGD-Nominal} for the fractal aggregates. The upper-left plot in Figs ~\ref{fig:Mie-Nominal}--~\ref{fig:RGD-Nominal} is the ratio
$F$(5 $\mu$m)/$F$(2 $\mu$m), where $F$ is the intensity of the signal, in flux or radiance units. In VIRTIS-H non-outburst spectra this ratio is 0.8--0.9, whereas values  of typically 0.4--0.5 characterise spectra of outburst material. Only models providing a good match to the observed bolometric albedo will be able to explain the relative intensity of the thermal emission with respect to scattered light.

   \begin{figure}
    \includegraphics[width=8.cm]{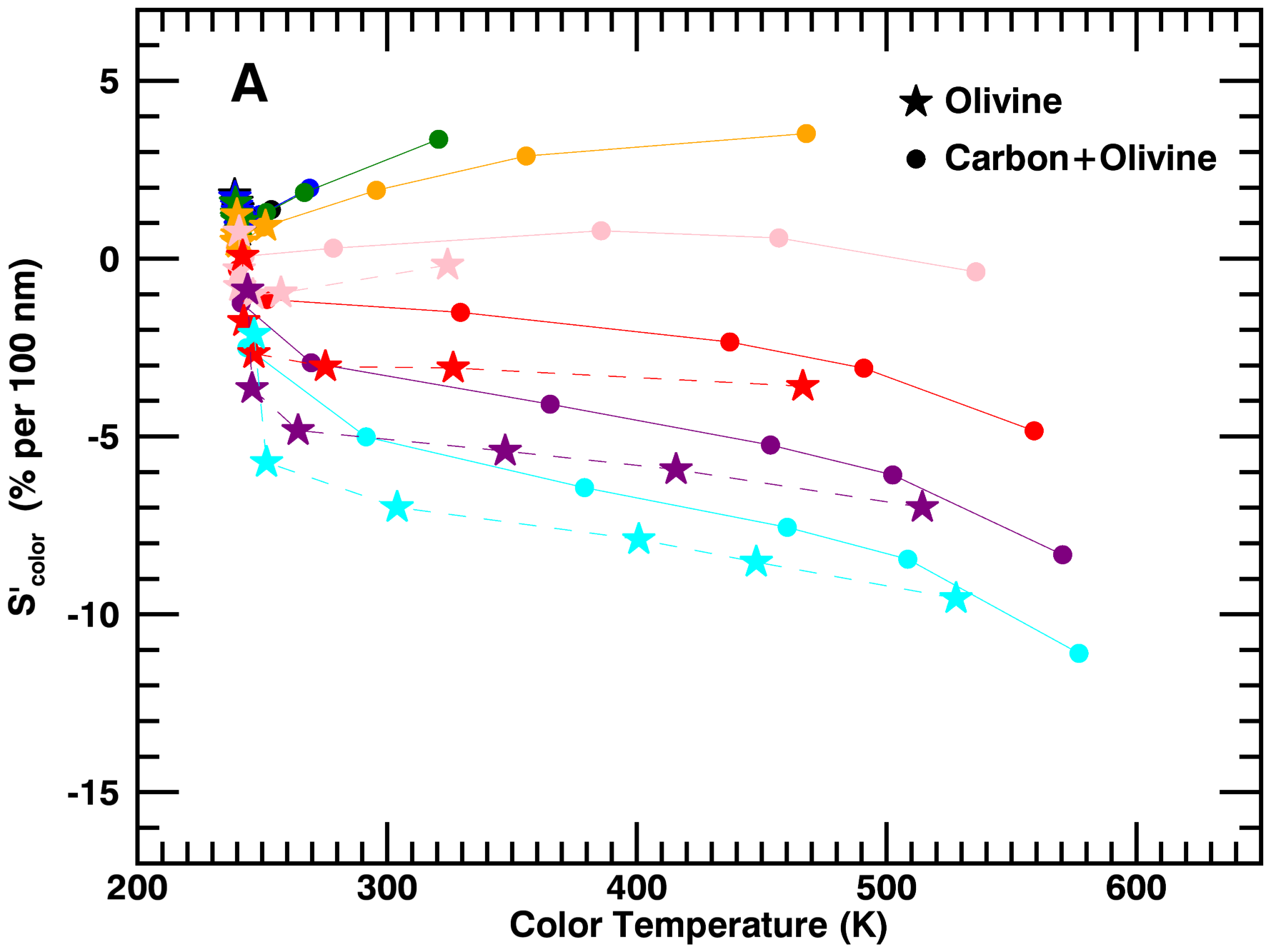}
    \includegraphics[width=8.cm]{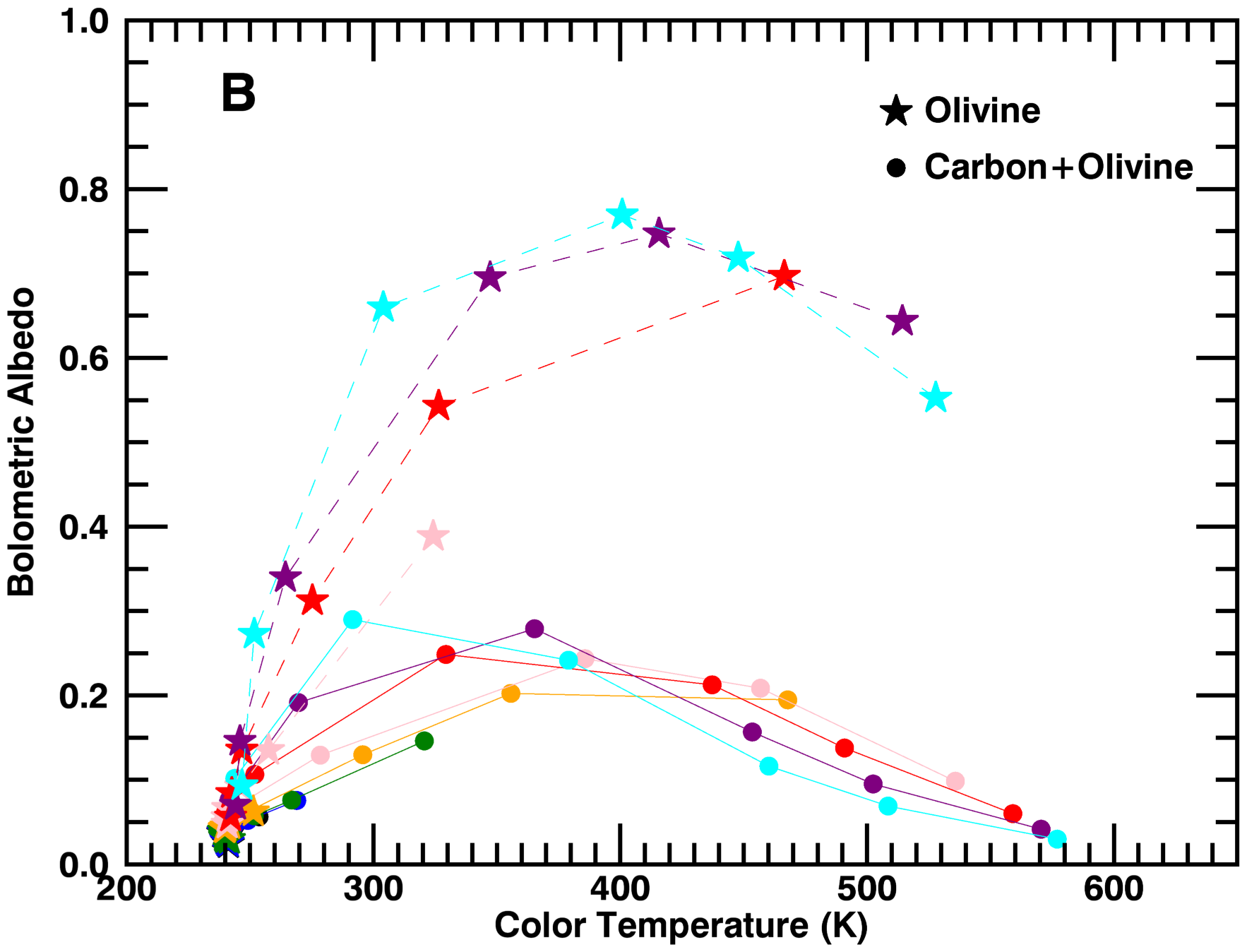}
    \includegraphics[width=8.cm]{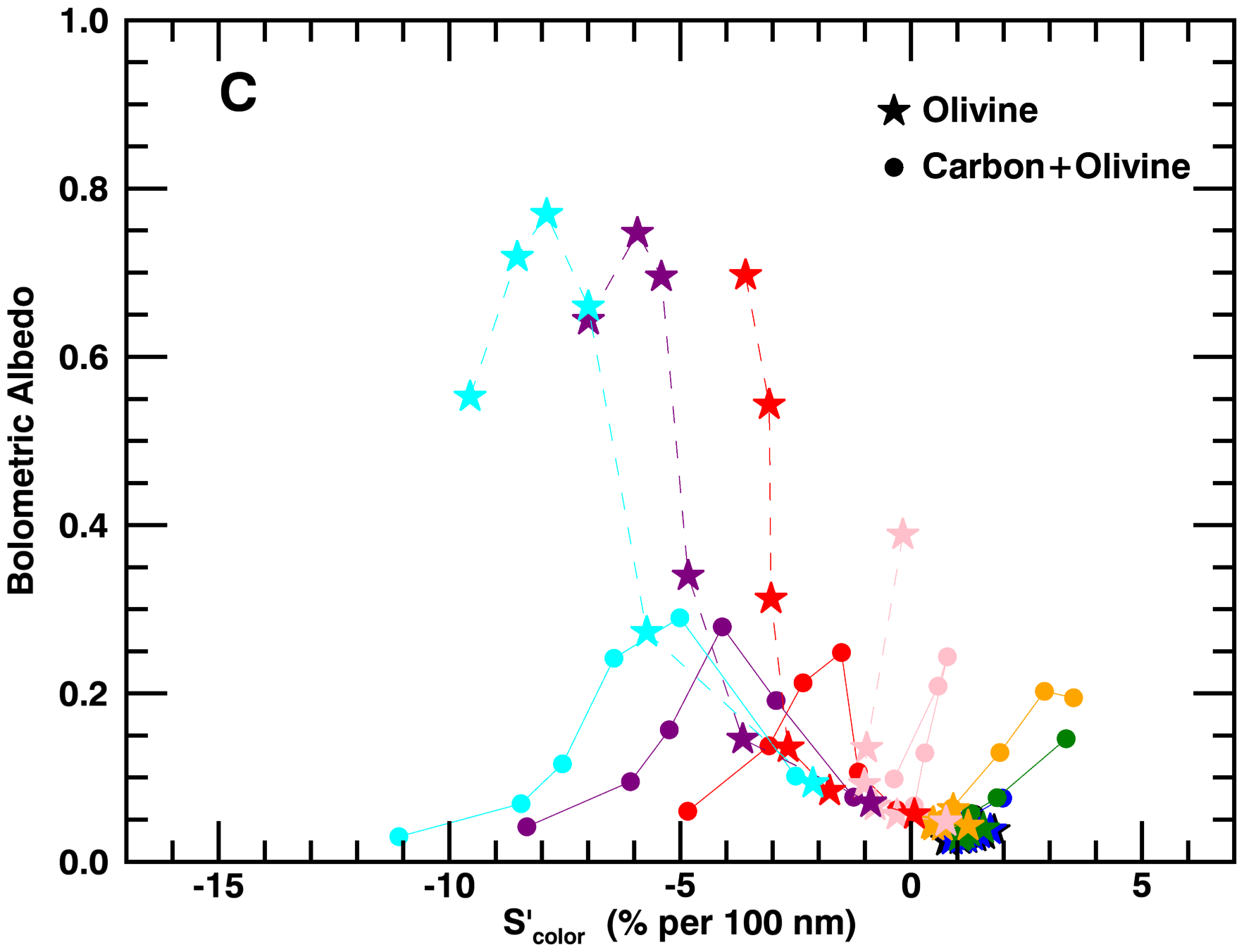}
    \caption{Results from Mie-scattering calculations: correlation between spectral properties. The different colours are for different size indices $\beta$ and correspond to those of Fig.\ref{fig:Mie-Nominal}. The different points for each $\beta$ refer to different minimum sizes $a_{\rm min}$. Dotted symbols connected with dashed lines are for a carbon/olivine porous mixture with $q_{\rm frac}$ = 0.7. Stars connected with dashed lines are for porous olivine grains. The Mg:Fe content of olivine is 50:50. The model parameters are $a_{\rm unit}$ = 100 nm, $D$ = 2.5, $P_{\max}$ = 0.5.}
    \label{fig:Mie-Correl}
\end{figure}

\subsection{Quiescent coma}
\label{sec:background}

The spectral properties of the 67P dust quiescent coma resemble those measured in other moderately active comets (Sect.\ref{sec:discussion}). As discussed in many papers in the literature, a 20\% colour temperature excess with respect to the equilibrium temperature ($S_{\rm heat}$ $\sim$ 1.2) could be explained by the presence of micron-sized and smaller grains comprising absorbing material. The red colour implies that small grains with respect to the wavelength are not dominant, otherwise blue colours would be observed.

Figure~\ref{fig:Mie-Nominal} shows that Mie scattering, for a moderately porous ($P_{\max}$ = 0.5) mixture of amorphous olivine and amorphous carbon, is able to reproduce the properties of the 67P dust coma, providing submicrometric particles are considered. The size distribution cannot be uniquely constrained. Indeed, the spectral properties depend both on the slope of the size distribution $\beta$ and minimum particle size $a_{\rm min}$. The sets ($a_{\rm min}$, $\beta$) = (0.2$\mu$m, 2.5) and (0.5$\mu$m, 3.0) provide satisfactory results for $T_{\rm col}$, $A(\theta)$, and $S'_{\rm col}$. Size distributions with $\beta \geq$ 3.5 are not able to reproduce the observed red colour. The same constraints are obtained considering mixtures of carbon with a somewhat lower content of silicates, with fully transparent silicates, and considering pure-carbon grains. For non porous grains ($P$=0), the colour temperature is matched with the  sets ($a_{\rm min}$, $\beta$) = (0.1$\mu$m, 2.5) and (0.3$\mu$m, 3.0), but $A(\theta$) is overestimated by a factor of two, whereas the colour is approximately matched.  Grains devoid of dark absorbing material are excluded, as shown in Fig.~\ref{fig:Mie-Correl}. This figure presents how the spectral properties vary with each other, for comparison with the observed correlation plots shown in Fig.~\ref{fig:correlation}.  It shows that, for pure silicatic grains or silicates mixed with FeS, size distributions consistent with the measured $T_{\rm col}$ yield neutral colours and bolometric albedos higher than measured.

\begin{figure}
   \includegraphics[width=8cm]{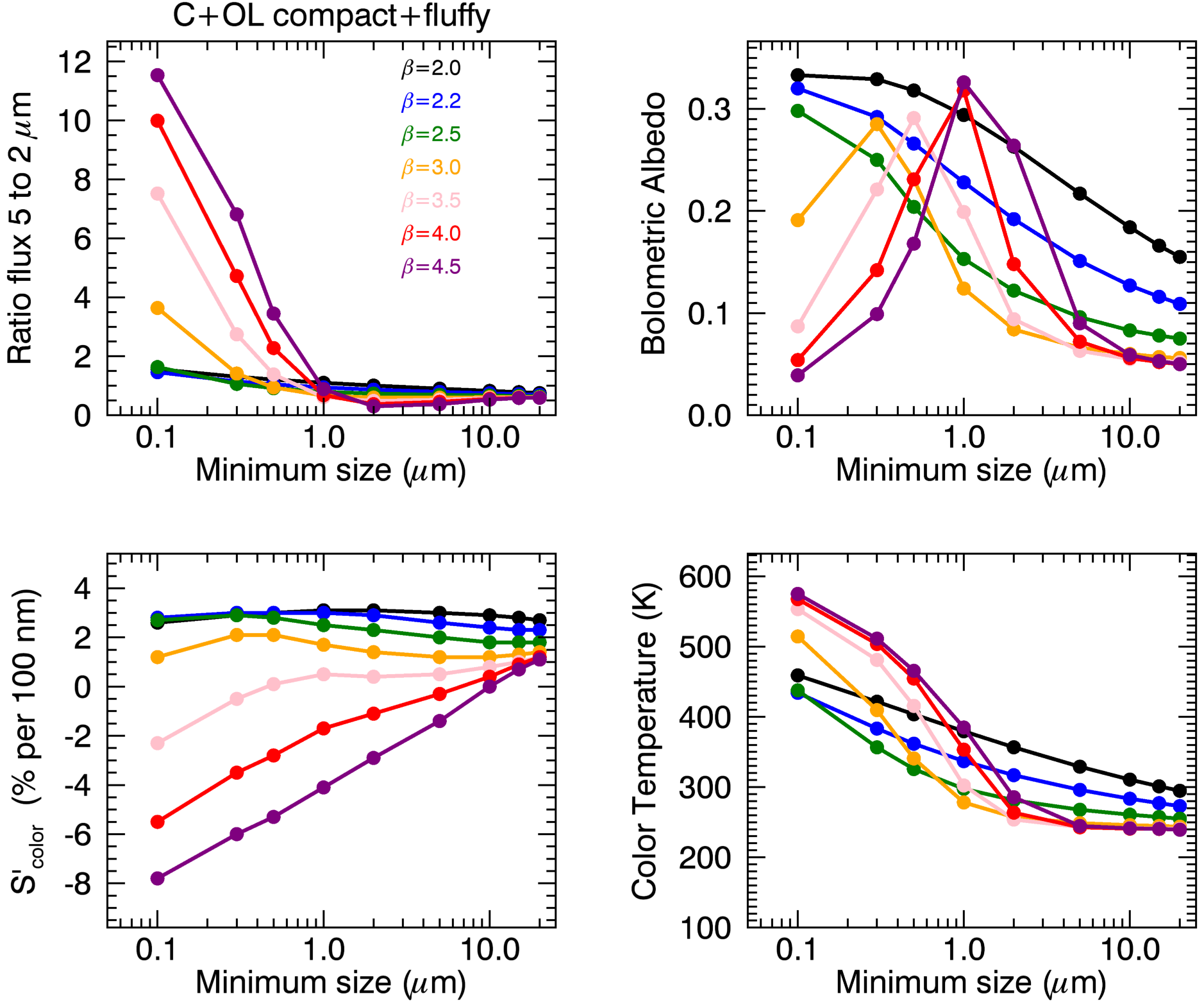}
    \caption{Results for relative fractions of 25\% of fractal aggregates and 75\% of moderately porous carbon/olivine mixtures ($P_{\max}$ = 0.5). See captions to Figs~\ref{fig:Mie-Nominal}--\ref{fig:RGD-Nominal}. The model parameters are: $q_{\rm frac}$ = 0.7, $a_{\rm unit}$ = 100 nm, $D$ = 2.5 for compact grains and $D$ = 1.7 for fluffy grains. The Mg:Fe content of olivine is 50:50.}
    \label{fig:fluffy}
   \end{figure}

The above interpretation does not consider fractal aggregates which are a significant component of the dust population in 67P coma, as discussed in Sect.~\ref{sec:model}. BCCA-type fractal aggregates reach high temperatures whatever their size (Fig.~\ref{fig:RGD-Nominal}), so they could contribute to the 20\% colour temperature excess. However, they are not expected to contribute significantly to the scattered emission, since their spectrum exhibits a low bolometric albedo  (Fig.~\ref{fig:RGD-Nominal}). \citet{Dellacorte2016} report the counts of compact particles compared with the number of fluffy fractal-like aggregates detected by GIADA from September 2014 to  February 2016. For the Sept. 2014--Feb. 2015 period (bound S/C orbits), the relative number of fluffy aggregates is $\sim$ 25\%. In the following months, this number was
$<$ 10\% (3\% at perihelion). The decrease  is  attributed to the low spacecraft potential at this time preventing charging and electrostatic fragmentation of the aggregates for detection with GIADA \citep[][Fulle, personal communication]{Fulle2015}.  We performed calculations with a dust population containing both types of particles assuming they have the same size distribution. A relative fraction of fractal agglomerates of 3\%
does not have significant impact on the dust continuum radiation. Figure~\ref{fig:fluffy} shows results for a relative number of fluffy aggregates of 25\%. The VIRTIS-H spectra can be explained providing the minimum radius of the compact particles exceeds $\sim$ 1 $\mu$m. For a minimum size
$a_{\rm min}$ = 1 $\mu$m, the required size index is $\beta$ = 2.5. For $a_{\rm min}$ = 15 $\mu$m, the model provides a good fit with $\beta$ = 2. Large size indices ($\geq$ 3.5) are excluded on the basis of the measured colour and bolometric albedo. Neither COSIMA nor MIDAS provide evidence for abundant micrometre-sized particles in the coma of 67P \citep{Merouane2016,Mannel2016}. Therefore, we can conclude that the VIRTIS-H spectra are consistent with the constraints obtained by the in situ dust instruments onboard Rosetta: a dust population consisting of a mixture of compact particles and fluffy particles with a shallow size index and particle sizes larger than a few microns.
In this picture, the dust superheating observed in comets is caused by fluffy aggregates. For 25\% of fluffy agregates in number, their relative contribution to the scattered light is less than 1.5\% at 2 $\mu$m.

   \begin{figure}
    \includegraphics[width=8.cm]{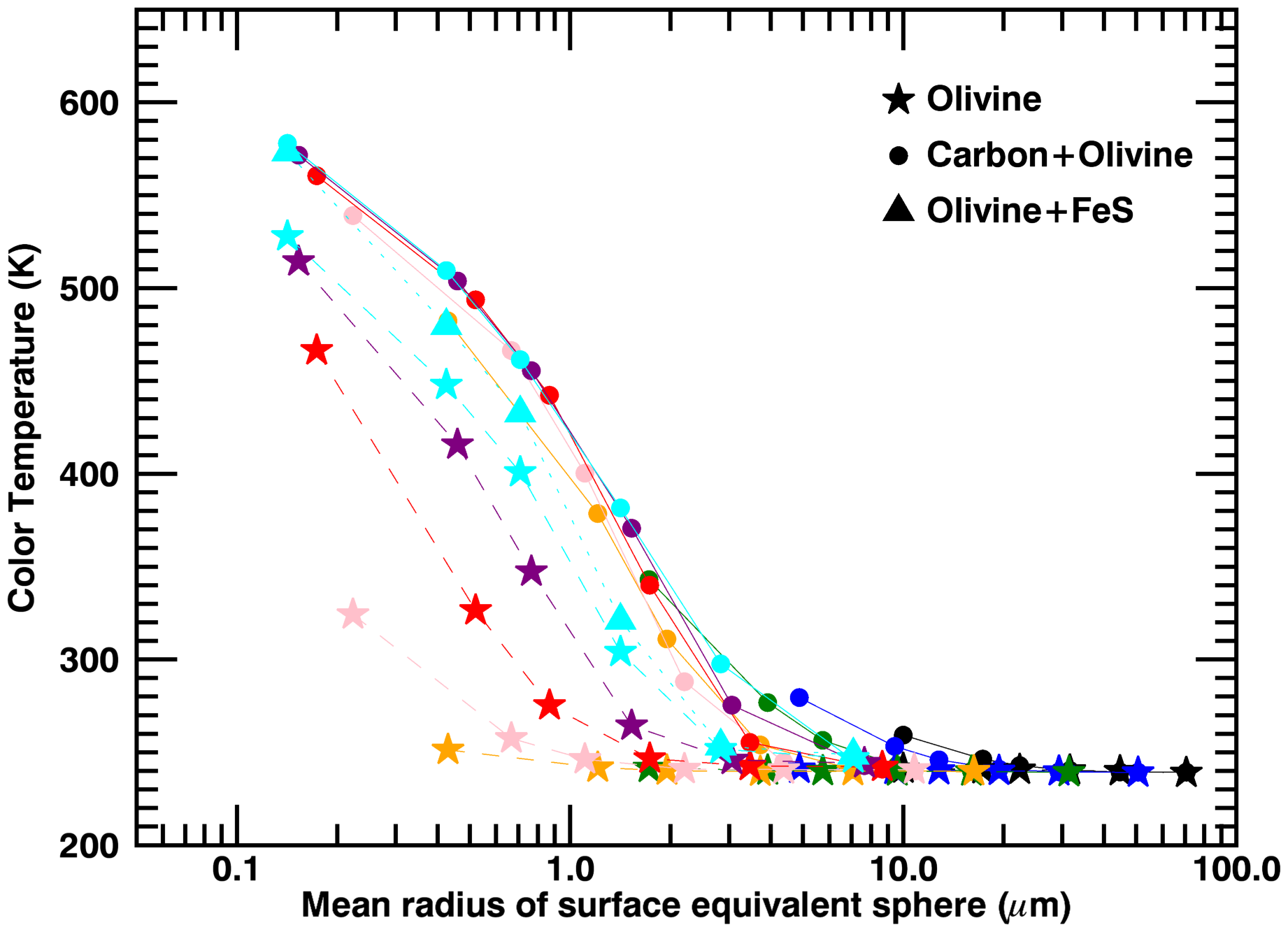}
    \caption{Results from Mie scattering calculations: colour temperature as a function of mean radius of surface equivalent sphere. Same colour coding as for Figs~\ref{fig:Mie-Nominal}--\ref{fig:Mie-Correl}. Filled triangles are for a porous mixture of olivine and FeS, with 6\% of FeS in volume (only results for $\beta$ = 5 are shown). }
    \label{fig:Mie-colT}
\end{figure}

\subsection{Outburst material}
\label{sec:outburst}
The IR continuum emission of the outburst material is characterised by high colour temperatures and bolometric albedo, and blue colours (Sect.~\ref{sec:properties}, Figs~\ref{fig:properties-13SEPT}--\ref{fig:properties-14SEPT}). The high colour temperatures and blue colours imply the presence of Rayleigh-type  scatterers in the ejecta, i.e., either very small grains or BCCA type agglomerates.

Let first consider the case of compact grains with $P_{\rm max}$ = 0.5. Figures~\ref{fig:Mie-Nominal} and \ref{fig:Mie-Correl} show that the high colour temperatures and blue colours are obtained in the case of large size indices $\beta$ and small $a_{\rm min}$. The colour temperature is plotted as a function of the mean radius of surface equivalent sphere in the dust population $a_{\rm mean}$ (calculated from the weighted mean of the particle cross-sections) in Fig.~\ref{fig:Mie-colT}. Colour temperatures of 600 K are reached  for carbon/olivine grains with a mean radius of $\sim$ 0.1 $\mu$m, whereas the colour temperature for silicatic grains of this size is $\sim$ 530 K. Silicatic grains reach a high temperature because we are here considering olivine with a high Fe content which is a significantly absorbing material. For the steepest size distribution ($\beta$ = 5), the colour reaches
values $\sim$ --10\% per 100 nm. Both the colour and colour temperature measured at the peak of brightness of the 14 Sept. outburst are satisfactorily reproduced with a  population of carbon/olivine grains, constituted almost uniquely of small grains with a mean radius of $a_{\rm mean}$ $\sim$ 0.1 $\mu$m. However, the high bolometric albedo $\sim$ 0.6 is not explained (Fig.~\ref{fig:Mie-Correl}B--C). The lower $T_{\rm col}$ and moderate bluish colour measured for the 13 Sept. outburst implies somewhat larger grains ($a_{\rm mean}$ $\sim$ 0.5 $\mu$m for carbon/olivine mixtures),
but again the high bolometric albedo excludes grains with a significant content of strongly absorbing carbonaceous compounds.

Figure~\ref{fig:Mie-Correl} shows that spectra of silicatic grains (with the optical properties of Fe:Mg amorphous olivine) exhibit high bolometric albedos when small particles are dominant. Furthermore, the correlations observed between the spectral properties for pure olivine (Fig.~\ref{fig:Mie-Correl}) present striking similarities with those observed in the course of the outbursts (Fig.~\ref{fig:correlation}). This
suggests that the outburst ejecta are comprising small and bright particles. The decrease of the colour temperature and colour shortly after the peak brightness suggests an increase of the mean particle size with time for the particles crossing VIRTIS FOV. The presence of silicatic grains devoid of organic material could be explained by the high grain temperatures causing sublimation of the organics.

Quantitatively, the 13 Sept. outburst data are consistent with  the silicatic grain hypothesis. However, large colour temperatures ($>$600 K) are measured at the peak brightness of the 14 Sept. outburst, whereas computed temperatures are $<$ 530 K for olivine grains. Several explanations are possible. First, we did not consider grains smaller than 100 nm in radius, whereas we cannot exclude their presence. These grains might reach higher temperatures than 100-nm grains, though this is not expected from model calculations \citep{Lasue2007}. In our model, 100-nm grains are compact ($P$ = 0). Considering instead that these grains are aggregates, their temperature would be higher than the temperature of the equivalent sphere \citep{Xing1997}. Another possibility is that the spectral indices of amorphous olivine differ significantly from those of the refractory material of cometary grains. We performed Mie calculations with inclusions of FeS  within an olivine matrix. For a volume fraction of 6\% of FeS consistent with the suggestion of \citet{Fulle2016} for 67P grains, the colour temperature reaches 575 K (Fig.~\ref{fig:Mie-colT}), which is not far from the extremum values measured during the 14 Sept. outburst.

Could fractal agglomerates explain the outburst spectra? For carbon/amorphous olivine  mixtures,  the colour and colour temperature of their infrared spectrum match the measurement, but this is not the case for the bolometric albedo (Fig.~\ref{fig:Mie-Correl}). Calculations made for units of amorphous olivine provide a bolometric albedo $<$ 0.4, i.e., a factor of two below the measured values. During 67P outbursts, MIDAS collected a high amount of micrometric grains  (Bentley, personal communication), in agreement with an ejecta population dominated by compact grains rather than aggregates.  We performed calculations setting the maximum size of the fractal aggregates to $a_{\max}$ = 1 $\mu$m, and the bolometric albedo is still $<$ 0.4.

Finally, the main contributors to the scattered and thermal parts of the spectrum could be distinct populations of grains. Cold and bright grains would essentially contribute to scattered light, whereas this is the opposite for warm and dark grains made of strongly absorbing material. These two populations of grains were
found in the outburst ejecta of the extraordinary outburst of comet 17P/Holmes in October 2007 ($r_{\rm h}$ = 2.45 AU) \citep{Yang2009}. The near-IR spectra of 
the outburst ejectas showed strong absorption bands at 2 and 3 $\mu$m related to water ice \citep{Yang2009}.

\begin{figure}
    \includegraphics[width=8.cm]{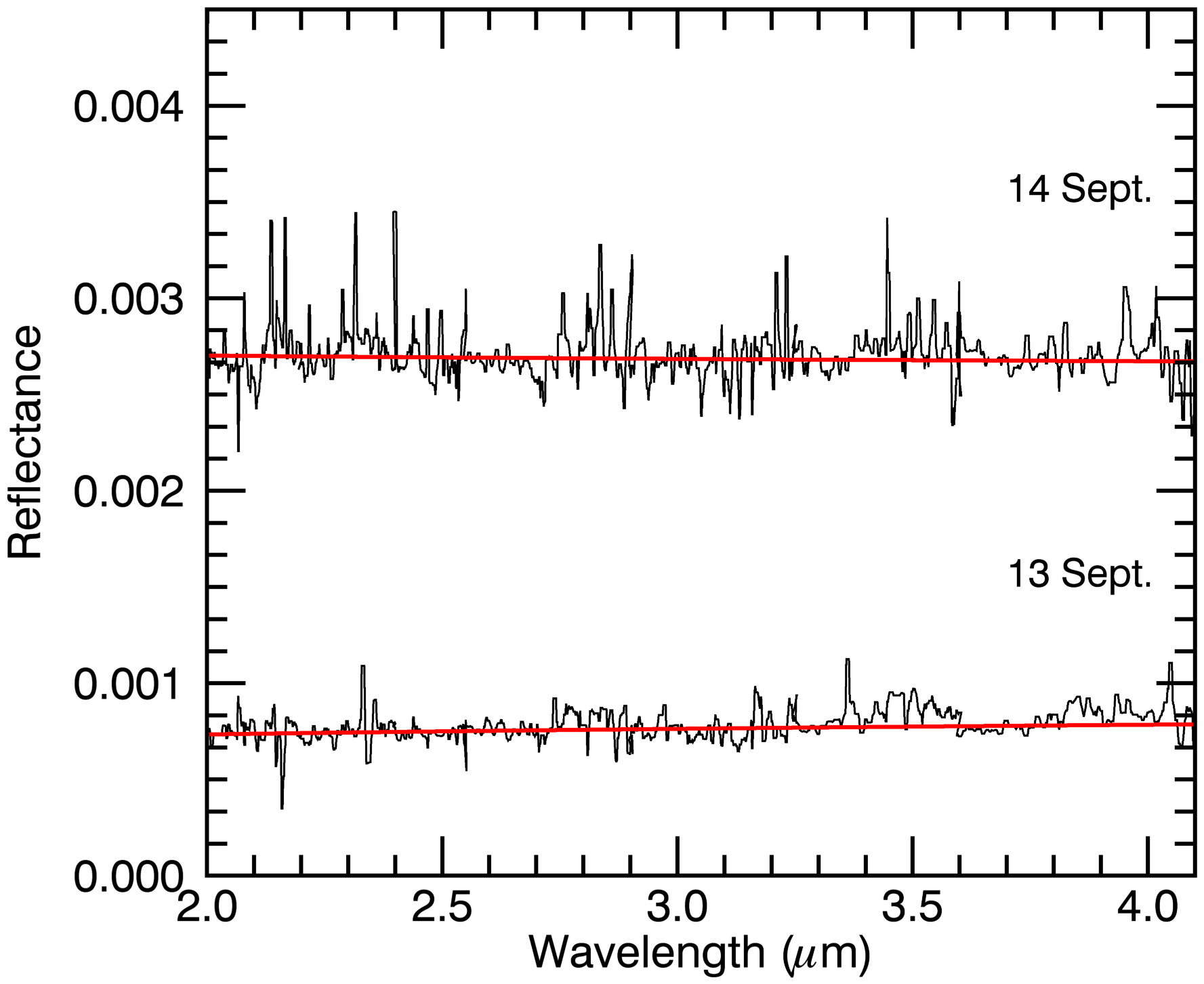}
    \caption{Reflectance spectra of outburst ejecta. Acquisitions 921--1051 (13 Sept. 13.71246--13.98618 h UT) and 661--691 (14 Sept.  18.90540--18.96821 h UT) were considered. The 14 Sept. spectrum has been shifted vertically by +0.002. A median smoothing with dimension 6 has been applied. The reflectance is defined by $R(\lambda)$= $\pi I(\lambda) r_{\rm h}^2 /F_{\odot}(\lambda)$, where $I(\lambda)$ is the measured radiance afer removing the thermal radiation. The solid red line is the fitted reflectance over the full VIRTIS-H spectrum. }
    \label{fig:reflectance}
\end{figure}

Infrared signatures of ice are not detected in the VIRTIS-H outburst spectra. The 2--$\mu$m ice signature extends up to 2.2 $\mu$m \citep[e.g.,][]{Yang2009}, but the sensitivity in this wavelength range is low. In the 2.7--3.5 $\mu$m range, the dust continuum includes a significant contribution of the thermal radiation ($\sim$ 50\% or more), because the colour temperature is high. Therefore, detection of the 3--$\mu$m ice band, as an absorption feature in the dust scattered emission, is difficult. Reflectance spectra, obtained after subtracting the thermal radiation, do not show any absorption band. Figure~\ref{fig:reflectance} shows outburst reflectance spectra averaging acquisitions in the time intervals 13 Sept. 13.71246--13.98618 h UT and 14 Sept. 18.90540--18.96821 h UT. The signal-to-noise ratio is 13--16 in the 3--$\mu$m ice band domain, implying a band depth less than $\sim$ 10\%.

\section{Spectral signatures: H$_2$O, CO$_2$ and organics}
\label{sec:gas}

  \begin{figure}
    \includegraphics[width=8.cm]{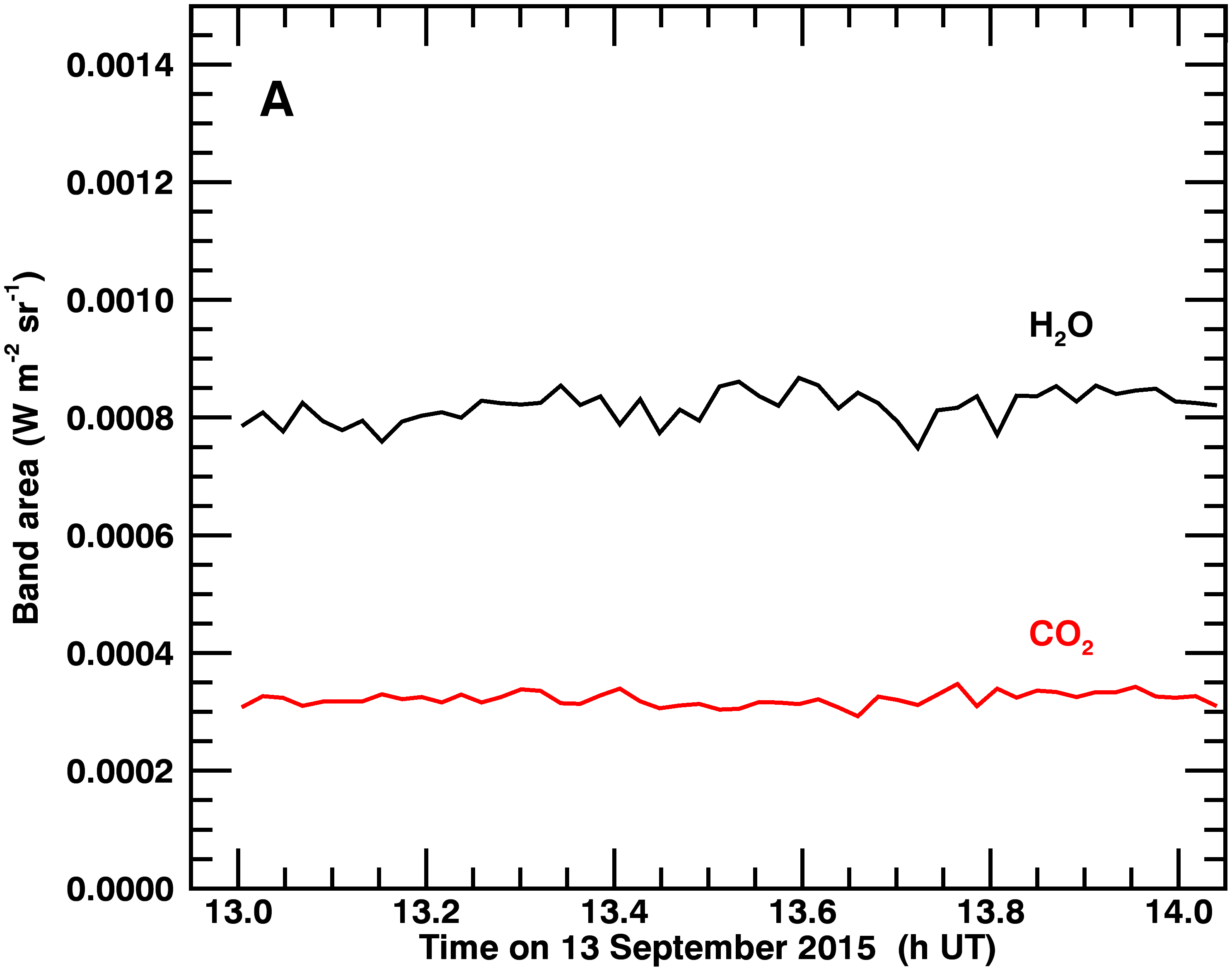}
           \par
\vspace{0.3cm}
    \includegraphics[width=8.cm]{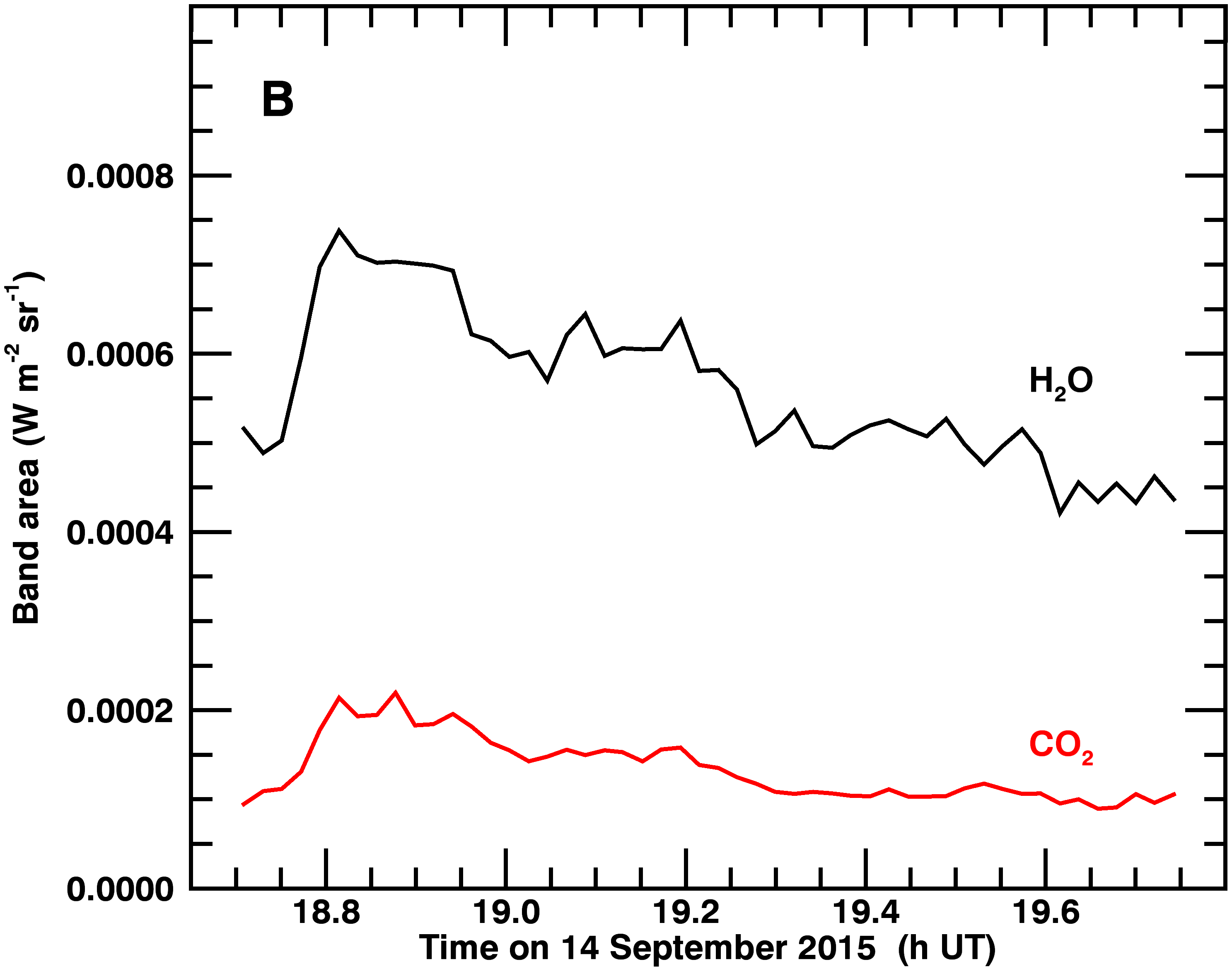}
           \par
\vspace{0.3cm}
    \caption{Temporal evolution of the areas of the H$_2$O 2.7 $\mu$m and CO$_2$ 4.26 $\mu$m bands at the time of the 13 Sept. (A) and 14 Sept. (B) outbursts. The time resolution is $\sim$ 70 s. The spectral ranges used for computing the band areas are 2.55--2.74 $\mu$m and 4.20--4.35 $\mu$m, for H$_2$O and CO$_2$  respectively. }
    \label{fig:mol-evolv}
\end{figure}

The VIRTIS-H spectra present strong H$_2$O and CO$_2$ fluorescence emission bands, at 2.67 and 4.26 $\mu$m, respectively (Figs~\ref{fig:spectra}--\ref{fig:spectra2}).
The temporal evolution of their band areas does not provide evidence for outburst-related variations (Fig.~\ref{fig:mol-evolv}). The variations observed for 14 September
in Fig.~\ref{fig:mol-evolv}B are mainly related to non-monotonic changes of the distance of the FOV to the comet centre during the time interval considered in the plot ($\rho$ from 9.28 to 4.83 km and then from 4.83 to 6.32 km, see Sect.~\ref{sec:obs}). Indeed, the band areas are found to be well correlated with $\rho$, varying as $\propto$ $\rho^{-0.7}$
and $\propto$ $\rho^{-1.17}$ for H$_2$O and CO$_2$, respectively. These variations show that, although the H$_2$O and CO$_2$ emissions are affected by optical depth effects \citep{dbm2016}, they respond to small changes in the column density. Observed 1--$\sigma$ fluctuations with respect to the mean band areas (for 13 Sept), and the mean band areas
corrected from distance (for 14 Sept.) are 3--5\% for H$_2$O and 4--12 \% for CO$_2$, the larger numbers referring to the 14 Sept. data. These numbers set an  upper limit to the contribution of a possible gaseous counterpart to these dusty outbursts. The determination of upper limits on the gas loss rates is beyond the scope of this paper.

  \begin{figure}
    \includegraphics[width=8.cm]{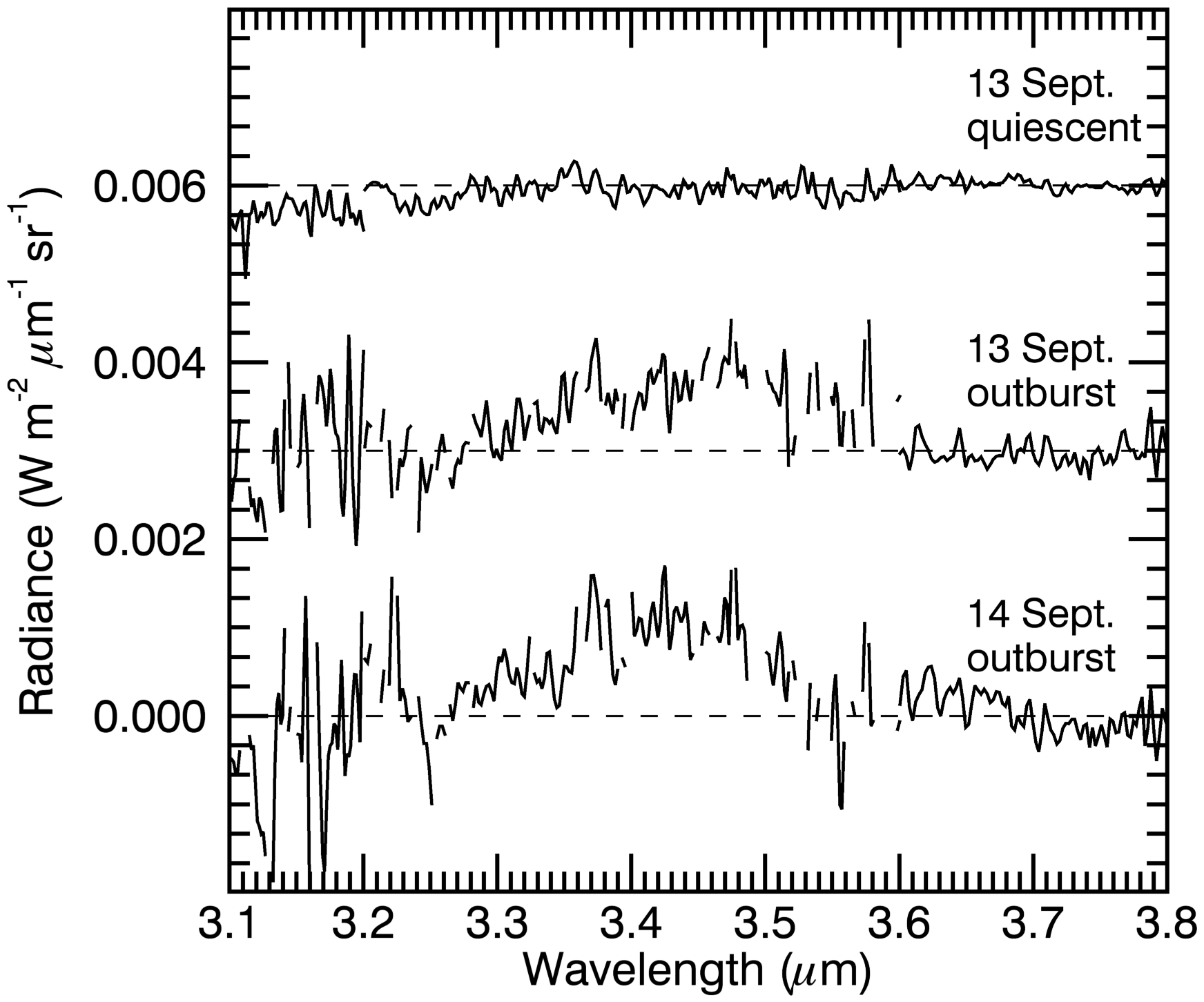}
    \caption{VIRTIS-H spectra of the 3.4-$\mu$m band observed on 13 and 14 Sept. 2015.  Labelled {\it outburst} spectra correspond to 13 Sept. 13.712--13.792 h UT (middle) and 14 Sept. 18.863-18.897 h UT (bottom). The spectrum labelled {\it quiescent} is the average of 850 acquisitions acquired before the 13 Sept. outburst.  The dust continuum emission has been removed. Outburst spectra include the contribution of the background coma, which is low. The {\it outburst} and {\it quiescent} 13 Sept. spectra have been shifted by +0.003 and +0.006, respectively, in radiance units. Spectral channels found to be much more noisy than nearby channels have been removed. }
    \label{fig:organics}
\end{figure}

A broad band extending between about 3.3--3.6 $\mu$m is detected in the outburst spectra (Fig.~\ref{fig:organics}). The strength of the band correlates with the strength of the continuum emission. The selected time ranges for the outburst spectra shown in Fig.~\ref{fig:organics} correspond to maximum band strength. The band is not detected in pre-outburst spectra (Fig.~\ref{fig:organics}). When averaging several VIRTIS-H data cubes acquired near perihelion, some intensity excess has been detected in this spectral region, especially from CH$_4$ lines \citep{dbm2016}.

This 3.3--3.6 $\mu$m band has been observed in several low-resolution cometary spectra \citep[][and references therein]{dbm1995}, in particular during the outburst of 17P/Holmes \citep{Yang2009}. It corresponds to the signature of C-H stretching modes from organic species and hydrocarbons. Cometary spectra obtained at high spectral resolution showed that ro-vibrational lines of C$_2$H$_6$ and CH$_3$OH, as well as a few lines of OH and H$_2$CO  \citep[e.g.][]{Dello2006}, and CH$_4$ \citep[e.g.,][]{Mumma1996}, are the main contributors to this broad band for comets observed in quiescent states.
Interestingly, the same strong line at 3.37 $\mu$m (2965--2968 cm$^{-1}$) is detected during the two 67P outbursts. Lines of C$_2$H$_6$ and CH$_3$OH fall within this range \citep{Dello2006}, but stronger lines from these species are not observed.

We postpone the analysis of this band to a future paper, when data with an improved calibration will be available. The VIRTIS data suggest abundant release of organics during 67P outbursts. In 9P/Tempel 1 Deep Impact ejecta, enhanced C$_2$H$_6$ infrared emission was observed, higher than for CH$_3$OH and H$_2$O lines  \citep{Mumma2005,DiSanti2007}.

\section{Summary and discussion}
\label{sec:discussion}

  \begin{figure}
    \includegraphics[width=8.cm]{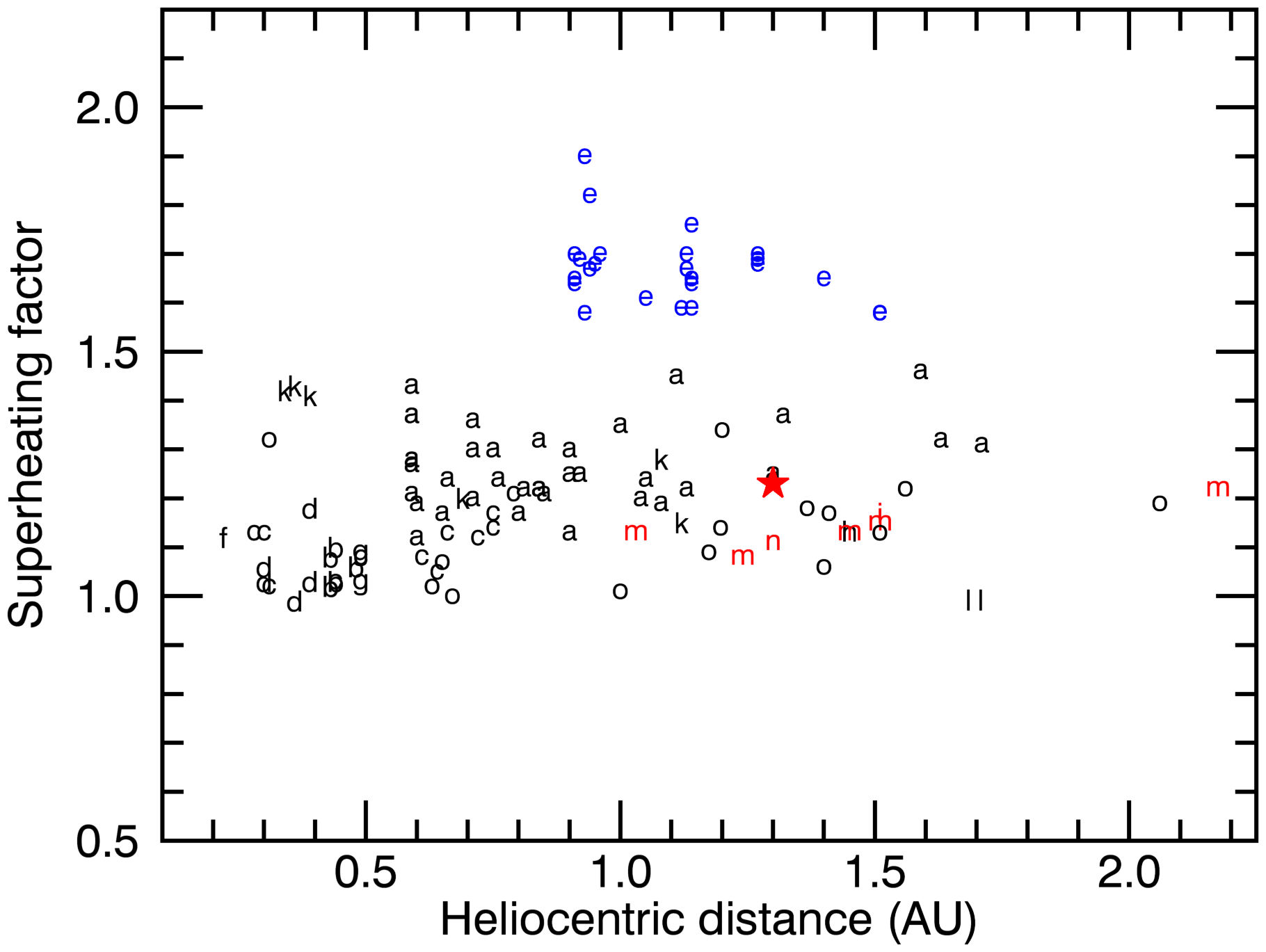}
    \includegraphics[width=8.cm]{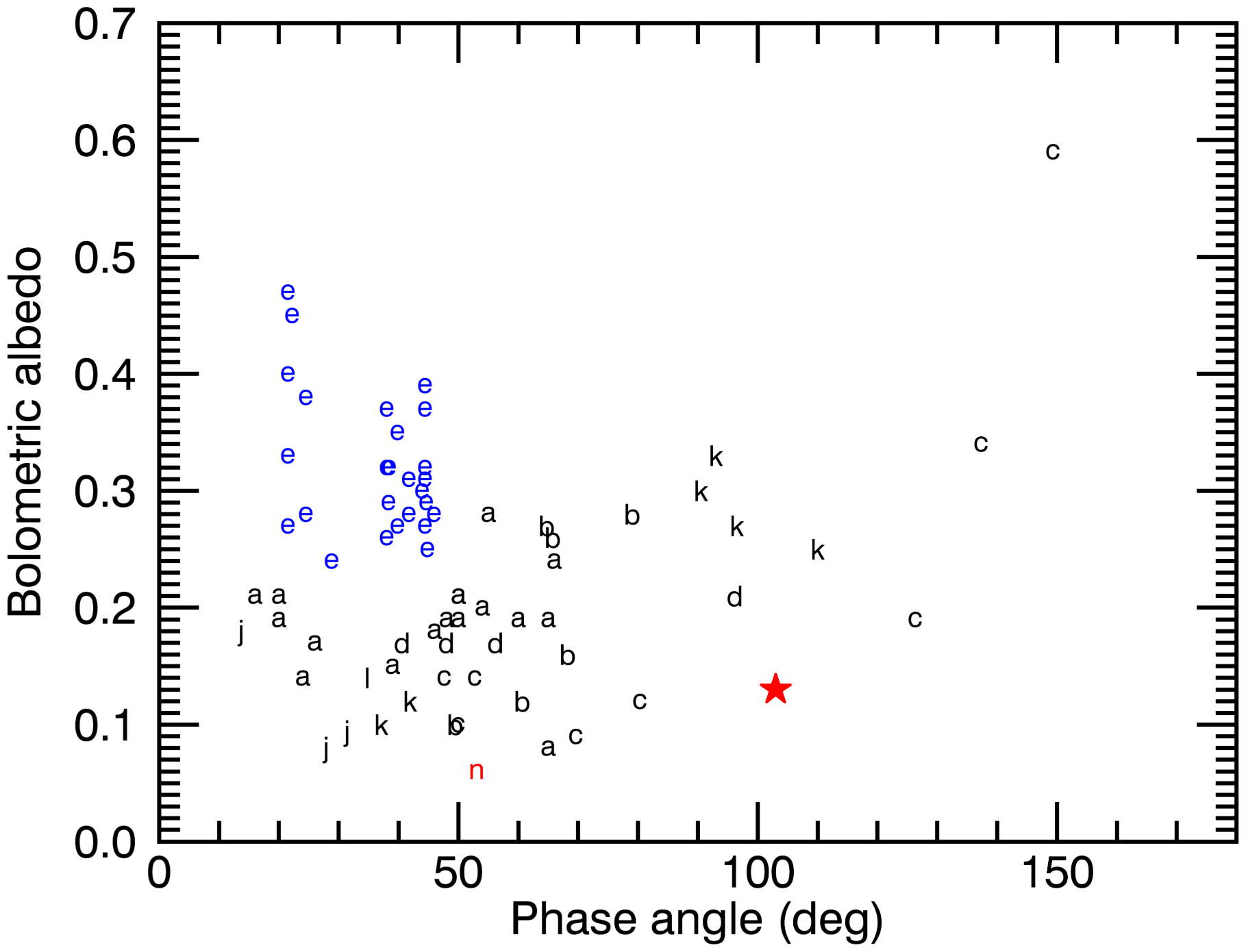}
    \caption{Superheat factor $S_{\rm heat}$ (top) and bolometric albedo $A$ (bottom) measured in comets (letters) compared to 67P values (red star) for the quiescent coma at 1.3 AU from the Sun (this work). Letters correspond to comets:
    (a) 1P/Halley, (b) C/1975 N1 (Kobayashi-Berger-Milon), (c) C/1980 Y1 (Bradfield), (d) C/1989 X1 (Austin), (e) C/1995 O1 (Hale-Bopp), (f) C/1985 K1 (Machholz), (g) 23P/Brorsen-Metcalf, (h) C/2007 N3 (Lulin), (i) 9P/Tempel 1, (j) C/2009 P1 (Garradd), (k) C/1996 B2 (Hyakutake), (l) C/2012 K1 (PAN-STARRS), (m) Jupiter-family comets 69P/Taylor, 4P/Faye, 19P/Borrelly, 55P/Tempel-Tuttle, 24P/Schaumasse, (n) 103P/Hartley 2, (o) Long-period comets reported in \citet{Sitko2004}. Other measurements are from \citet{Gehrz1992,Gicquel2012,Mason1998,Mason2001,Meech2011,Sitko2013,Woodward2011,
Protopapa2014,Woodward2015}. Blue colours are used for comet Hale-Bopp. A red colour is used for Jupiter-family comets.}
    \label{fig:comp-comets}
\end{figure}

We have presented an analysis of the 2--5 $\mu$m dust continuum
radiation from comet 67P measured at 1.3 AU from the Sun with the
VIRTIS-H instrument. We focused on data acquired on 13 and 14
September 2015 at the time of two important 67P outbursts, with the
aim to characterise the properties of the dust ejecta, in addition
to those of the quiescent coma. A discussion of
the results obtained in this paper is given below.

{\it Comparison of the quiescent coma of 67P with other comets.} The coma of 67P at 1.3 AU from the Sun observed at a phase angle of $\sim$100$^\circ$ exhibits a bolometric albedo of $\sim$0.13, a colour of $\sim$ 2.5\% per 100 nm in the range 2--2.5 $\mu$m, and a colour temperature which is 20\% in excess with respect to the equilibrium temperature. Figure~\ref{fig:comp-comets} shows the bolometric albedo and superheat for a sample of comets. Within a few exceptions, 67P dust properties are in the mean of values measured in other comets, in particular Jupiter-family comets. Higher values for the bolometric albedo and superheat are measured for the bright comet C/1995 O1 (Hale-Bopp) \citep{Mason2001}, whereas the hyperactive comet 103P/Hartley 2 shows an unsually low bolometric albedo of 0.06 at 53$^{\circ}$ phase angle \citep{Meech2011} and a moderate superheat of 1.11 \citep{Protopapa2014}. The 2--2.5 $\mu$m colour measured in 67P is also typical \citep{Jewitt1986}. We measured a value of 0.02 for the product of the geometric albedo times the phase function, consistent with previous values measured for 67P \citep{Hanner1985} and with measurements in other comets \citep{Hanner1989}. Based on available data on the phase function \citet{Schleicher2010}, the geometric albedo of the quiescent 67P dust coma is estimated to $A_{\rm p}$ = 0.04. Altogether, the scattering and thermal properties of the quiescent dust coma of 67P are representative of the properties measured for moderately active comets.

{\it Dust properties in 67P quiescent coma.} Mie scattering modelling shows that the measured scattering and thermal properties imply the presence of hot submicrometre-sized  particles composed of absorbing material. This conclusion is consistent with previous interpretations of dust optical and infrared data \citep[e.g.,][and references therein]{Hanner2003,Kolokolova2004}. Fractal-like aggregates have been detected in significant quantities by the GIADA and MIDAS dust experiments onboard Rosetta, together with moderately porous compact grains \citep{Fulle2015,Bentley2016,Mannel2016}. Fractal aggregates with submicrometric absorbing subunits reach high temperatures, whereas they are not efficient scatterers. We showed numerically that such fluffy aggregates could be at the origin of the superheat, with the implication that submicrometric compact grains may be absent in cometary atmospheres.

{\it Size distribution in 67P quiescent coma.}  We have
obtained some constraints on the size index in 67P quiescent coma
at 1.3 AU from the Sun, having in mind model approximations and
assumptions. Overall, assuming that compact grains are responsible
for the superheat, a size index in the range 2.5--3 provides
satisfactory results for both the superheat, colour and bolometric
albedo. The low value $\beta$ = 2.5 is obtained for a minimum
particle-radius $a_{\rm min}$ of 0.2 $\mu$m, whereas $\beta$ = 3
is obtained for $a_{\rm min}$ = 0.5 $\mu$m. A somewhat shallower
size index, in the range 2--2.5, is deduced when considering a
mixture of compact particles and fractal aggregates, with ratio
75:25 in number. GIADA collections combined to OSIRIS data
obtained at 2.0 AU pre-perihelion are consistent with a size index
of 2.2--2.6 ($\pm$ 0.3) in the size range 80 $\mu$m to 2 mm
\citep{Rinaldi2017b}. GIADA collections for the 3.4--3.6 AU
pre-perihelion range show a similar size index \citep[$\sim$
2,][]{Rotundi2015}. A power index of 1.9 $\pm$ 0.3 is deduced from
the COSIMA collections in the size range 30--150~$\mu$m
\citep{Merouane2016}. However, \citet{Fulle2016b} concluded to a
steeper size index of 3.7 at perihelion from GIADA data, and found
that the heliocentric evolution of the size index is consistent
with ground-based observations of 67P dust tail and coma. The
VIRTIS-H data seem to exclude such a steep size distribution at
perihelion: both in the case of compact particles and
compact/aggregates mixtures, the model calculations with $\beta$ =
3.5--4 provide neutral or slightly blue colours for the $a_{\rm
min}$ values which provide colour temperatures consistent with the
observed value (Figs~\ref{fig:Mie-Nominal}, \ref{fig:fluffy}).
Therefore, though an evolution of the size distribution with
heliocentric distance is possible, VIRTIS data suggest that it is
not as strong as indicated by \citet{Fulle2016b}.

{\it Dust properties in 67P outburst ejecta.} 67P underwent many
dusty outbursts near perihelion \citep{Vincent2016b}. VIRTIS provides unique
information on the properties of the dust ejecta. The outburst material displays high colour temperatures (up to 630 K)
and blue colours (down to --10\% per 100 nm) in the IR. The
outburst plume is also less red at 0.55 $\mu$m than the quiescent
coma \citep{Rinaldi2017}. The two former properties can be
explained by the presence of very small grains. The constantly
high ($\sim$ 0.7) bolometric albedo measured during the outburst
sequence signs the presence of bright material in the outburst
ejecta. This bright material could be silicatic dust devoid of
organic material, or, alternatively, icy grains. The first
hypothesis would mean that the warm temperatures experienced by
the grains caused rapid degradation of the organic material. Insoluble organic
matter (IOM) from meteorites, which shares commonalities with 67P
refractory organics \citep{Fray2016}, disappears within $\sim$90 s
at 800 K and within 200 yr at 370 K \citep{Kebukawa2010}, from
which we derive that the time needed for thermal degradation at
600 K is $\sim$ 3 h. This is too slow to alter 67P grains observed
a few minutes after outburst onset. In presence of minerals, the
thermal decomposition rate of IOM is observed to be larger, but
the catalysing effect of minerals is efficient only for
temperatures less than 600 K \citep{Kebukawa2010}.

The second interpretation would be consistent with the detection of the signature of icy particles with the Alice UV spectrometer during dusty outbursts \citep{Steffl2016,Agarwal2017}, though the infrared signature of water ice at 3 $\mu$m is not detected in VIRTIS-H spectra. In addition, studies of morphological changes associated to dusty outbursts showed that these outbursts result from cliff collapse \citep{Pajola2017}. Images of the collapsed edge of the Aswan cliff revealed the fresh icy interior of the comet, with an albedo at least six times higher than the nucleus surface \citep{Pajola2017}. Therefore, the presence of icy grains in the dust plume is not unexpected. 

The very fine dust particles detected by VIRTIS in the outburst ejectas 
are possibly subunits of aggregates which fragmented or were superficially eroded during cliff collapse. Whether this is viable explanation needs quantitative investigations which are beyond the scope of this paper. The boulder size distribution in the Aswan talus, with a size index $\beta$ = 3.61, shows that collapsing blocks on 67P produce predominantly small chunks \citep{Pajola2017}, as observed for rock avalanche deposits on Earth.   Alternatively,  it is 
also conceivable that these small grains are present in surface and subsurface layers.
We have indeed some evidence for the presence of charged nanograins in the coma of 67P  \citep{Burch2015}. Under normal activity, the release of sub-micrometer-sized particles  is expected to be inefficient on cometary surfaces because the low gas pressure cannot overcome the relatively high tensile strength of small-size aggregate layers \citep{Gundlach2015}. Gas loading of these small particles become possible if they are released into the atmosphere during the cliff collapse.

{\it Mass loss in 67P outbursts.} Presently published estimations of the dust mass emitted in 67P outbursts are mainly derived from optical images obtained less than a few minutes after outburst onset \citep{Knollenberg2016,Vincent2016b}. These estimations make the assumption of a dust size distribution and geometric albedo similar to the quiescent coma values. However, the burst of brightness observed in scattered light is essentially from small and bright dust particles, as shown in this paper. Re-estimations in light of this finding have to be done.

{\it Gaseous counterparts in 67P dusty outbursts.} VIRTIS did not detect any increase in the H$_2$O and CO$_2$ column densities during the 13--14 Sept. outbursts. This means that the trigerring mechanism is not related to a large internal gas pressure, a mechanism proposed by \citet{Prialnik1993} for explosive outbursts. As already discussed, these outbursts were very likely caused by cliff collapse as for the one studied by  \citet{Pajola2017}. The lifetime of 0.1--1 $\mu$m icy grains is $\sim$ 300--800 s for dirty grains, and $>$ 10$^4$ s for pure ice grains \citep{Beer2006,Gicquel2012}. These long lifetimes may be one of the reasons why H$_2$O vapour was not detected in the VIRTIS-H outburst spectra. On the other hand, abundant release of organics is suggested by the detection of the 3.4 $\mu$m band in outburst spectra, possibly related to the thermal decomposition of complex organics present in grains, or to volatile organics originating from the collapsed wall.

{\it 67P outburst properties: comparison with other comets.} One of the best studied outbursts is the huge outburst of comet 17P/Holmes at 2.4 AU from the Sun. The ejecta displayed both a large superheat ($S_{\rm heat}$ = 2.0) and a blue spectral spectral slope in the IR (--3\% per 100 nm) \citep{Yang2009} which showed temporal variations attributed to changes in particle size. In addition, \citet{Ishiguro2010} measured a large geometric albedo two days after the outburst. The 2 and 3 $\mu$m water ice signatures were detected, with a depth for the 3-$\mu$m band of $\sim$ 30\%, i.e., much larger than the limit we obtained for 67P. A broad emission feature was also detected at 3.36 $\mu$m \citep{Yang2009}. Hence, there is strong similarity between 67P and 17P outburst ejecta. Other outbursting or fragmenting comets where an increase of superheat, bolometric albedo or blue colours have been measured include 1P/Halley, C/1996 B2 (Hyakutake), C/1999 S4 (LINEAR) \citep{Gehrz1995,Mason1998,Bonev2002}. We also note  similarities with the ejecta cloud produced by the Deep Impact experiment onto 9P/Tempel 1 composed of both icy grains and hot small-sized dust  \citep[e.g.,][]{Sunshine2007,Gicquel2012}.

To conclude, infrared observations with VIRTIS are providing unique information on the dust properties in 67P quiescent coma and outburst plumes, which are complementary to those from in situ and other remote sensing instruments onboard Rosetta. Future studies will benefit from coordinated multi-instrument data analyses.

\section*{Acknowledgements}
The authors would like to thank the following institutions and
agencies,  which supported this work: Italian Space Agency (ASI -
Italy), Centre National d'Etudes Spatiales (CNES- France),
Deutsches Zentrum f\"{u}r Luft- und Raumfahrt (DLR-Germany),
National Aeronautic and Space Administration (NASA-USA). VIRTIS
was built by a consortium from Italy, France and Germany, under
the scientific responsibility of the Istituto di Astrofisica e
Planetologia Spaziali of INAF, Rome (IT), which lead also the
scientific operations.  The VIRTIS instrument development for ESA
has been funded and managed by ASI, with contributions from
Observatoire de Meudon financed by CNES and from DLR. The
instrument industrial prime contractor was former Officine
Galileo, now Selex ES (Finmeccanica Group) in Campi Bisenzio,
Florence, IT. The authors wish to thank the Rosetta Science Ground
Segment and the Rosetta Mission Operations Centre for their
fantastic support throughout the early phases of the mission. The
VIRTIS calibrated data shall be available through the ESA's
Planetary Science Archive (PSA) Web site. With fond memories of
Angioletta Coradini, conceiver of the VIRTIS instrument, our
leader and friend. D.B.M. thanks L. Kolokolova for useful discussions.


\appendix
\section{Scattered and thermal emission}
\label{app:emission}

The scattered solar flux and thermal emission from a population of dust particles are respectively given by:

\begin{equation}
F_{\rm scatt}(\lambda) = K \frac{F_{\odot}(\lambda)}{4 \pi r_{\rm h}^2} \int_{\rm a_{\rm min}}^{{\rm a_{\rm max}}} C_{\rm scatt}(a,\lambda) p(a,\lambda,\theta') n(a)da,
\end{equation}

\begin{equation}\label{eq:A2}
F_{\rm therm}(\lambda) = K \int_{\rm a_{\rm min}}^{{\rm a_{\rm max}}} B(\lambda,T(a)) C_{\rm abs}(a,\lambda) n(a)da,
\end{equation}

\noindent
with :

\begin{equation}
B(\lambda,T) = \frac{2 h c^2}{\lambda ^5} \frac{1}{e^{h c/\lambda k T}-1}.
\end{equation}

\noindent
The scattering and absorption cross-sections are related to the scattering and absorption efficiencies $Q_{\rm scatt}(a,\lambda)$ and $Q_{\rm scatt}(a,\lambda)$:

\begin{equation}
C_{\rm scatt}(a,\lambda) = Q_{\rm scatt}(a,\lambda)G(a),
\end{equation}

\begin{equation}
C_{\rm abs}(a,\lambda) = Q_{\rm abs}(a,\lambda)G(a).
\end{equation}

In these equations, $a$ is the particle radius (from $a_{\rm min}$ to $a_{\rm max}$), $n(a)$ = $a^{-\beta}$ is the size distribution, $T(a)$ is the grain temperature, $G(a)$ = $\pi a^2$ is the particle cross-section. $K$ is a constant which sets the number of grains in the FOV. The solar irradiance $F_{\odot}$ at $r_{\rm h}$ = 1 AU from the Sun is modelled by a blackbody at 5770 K. The quantity $p(a,\lambda,\theta')$
is the phase function, where $\theta'$ is the scattering angle, which satisfies the normalization condition:

\begin{equation}
\frac{1}{2} \int_{0}^{\pi}p(a,\lambda,\theta') sin(\theta') d\theta' = 1.
\end{equation}

At thermal equilibrium, the temperature $T(a)$ can be computed from the balance of the solar energy absorbed at visual wavelengths and the energy radiated in the infrared \citep{Kolokolova2004}:

\begin{equation}\label{eq:tdust}
\frac{1}{r_{\rm h}^2} \int F_{\odot}(\lambda) C_{\rm abs}(a,\lambda) d\lambda = 4 \pi \int B(\lambda,T(a)) C_{\rm abs}(a,\lambda) d\lambda.
\end{equation}


Mie-scattering theory provides $Q_{\rm scatt}(a,\lambda)$, $Q_{\rm abs}(a,\lambda)$, and $p(a,\lambda,\theta')$
scattering parameters for homogeneous spheres made of material with refractive index $m$.

\section{Rayleigh-Gan-Debye theory}

The Rayleigh-Gan-Debye (RGD) theory is a simple and robust method to compute the scattering and thermal emission of fractal dust aggregates where the composent subunits are  Rayleigh scatterers \citep{Tazaki2016}. This theory has been thoroughly investigated
in the field of atmospheric science, to model, e.g., the optical properties of soot particles. It assumes that multiple scattering inside the aggregates can be ignored, so that the light scattered by the subunits are superposed, taking into account the phase differences between light rays. The RGD theory, its conditions of validity, and comparisons with results obtained with the more rigourous T-matrix method (TMM)  are presented in details in \citet{Tazaki2016}.
When the conditions of validity are satisfied, then the phase matrix element $S_{11}^{agg}$ is related to that of the subunits $S_{11}^{unit}$ through:

\begin{equation}\label{eq:rgd}
S_{11}^{\rm agg}(\theta') = N^2 S_{11}^{\rm unit}(\theta') S(q)
 \end{equation}

\noindent
where $N$ is the number of subunits and $S(q)$ is a structure factor, with $q = (4\pi/\lambda)sin(\theta'/2)$.

The phase matrix element $S_{11}$ is related to the scattering efficency and phase function through:

\begin{equation}\label{eq:S11}
S_{11}(\theta') = \frac{\pi}{\lambda^2} G(a) Q_{\rm scatt}(a,\lambda) p(a,\lambda,\theta').
\end{equation}

The structure factor $S(q)$ is a function of the fractal dimension $D$, the number of monomers, and a factor which describes the degree of correlation between the light rays scattered by the subunits. We used the expression $S(q)$ given by Eq. 24
of \citet{Tazaki2016}.

The number of subunits is given by:

\begin{equation}
N = k_0 (\frac{a}{a_{\rm unit}})^D,
\end{equation}

\noindent
where the size of the aggregate $a$ corresponds here to the radius of gyration $R_g$ (equal to (3/5)$^{ 1/2}$ times the characteristic radius for spherical aggregates). Calculations were made for  $D$ = 1.7 and $a_{\rm unit}$ = 100 nm. BCCA aggregates tend to have $D$ = 1.9--2
and $k_0$ = 1.03, so we adopted this value for $k_0$. $S(q)$ has the limits $S(q)$ = 1 for $qR_g << 1$ and $S(q)$ = $(qR_g)^{-D}$ for $qR_g >> 1$.

The angle-averaged scattering cross-section is given by \citep{Dobbins1991}:

\begin{equation}
C_{scatt}^{\rm agg} = N^2 G(a_{\rm unit}) Q_{\rm scatt}(a_{\rm unit},\lambda) \Big[1+\frac{4}{3D}\Big(\frac{2\pi R_g}{\lambda}\Big)^2\Big]^{-D/2}.
 \end{equation}

The absorption cross-section (that is the absorption efficiency times the aggregate cross-section) is equal to the total geometric cross-section of the subunits multiplied by the absorption efficiency for a single subunit \citep{Dobbins1991}:

\begin{equation}
C_{abs}^{\rm agg} = N G(a_{\rm unit}) Q_{\rm abs}(a_{\rm unit},\lambda).
 \end{equation}

Therefore, the temperature of the fractal aggregates is equal to the temperature of the subunits which can be computed from Eq.~\ref{eq:tdust}.

The criteria of validity for the RGD method are $\vert m-1\vert << 1$, 2$x_{\rm unit}\vert m-1 \vert << 1$, and 2$x_{\rm agg}\vert m_{\rm eff}-1\vert << 1$, where $x_{\rm unit}$ and $x_{\rm agg}$ are the size parameters of the subunits and aggregates, respectively \citep{Tazaki2016}. $m$ is the refractive index of the subunits. $m_{\rm eff}$ is the effective refractive index of the aggregates computed with the Maxwell-Garnett formula with the vacuum taken as the matrix component \citep{Greenberg1990}. Those criteria show that the RGD method provides the most reliable results for almost transparent monomers.
In our computations with amorphous carbon mixed with silicates the three quantities
$\vert m-1 \vert$, 2$x_{\rm unit}\vert m-1 \vert$, and 2$x_{\rm agg} \vert m_{\rm eff}-1 \vert$ are $\sim$ 1.1, 0.7, and 0.6, respectively, at $\lambda$ = 2$\mu$m. Therefore the validity conditions are only marginally satisfied for this opaque mixture. The comparison with TMM calculations made by \citet{Tazaki2016} suggests that we should expect reliable results within 10 \% or so for the scattered emission as far as BCCA particles are concerned.

\citet{Tazaki2016} showed that the RGD theory agrees with the TMM within 10\% for aggregates with $N$ $<$ 1024, which corresponds to an aggregate characteristic radius $\sim$ 4 $\mu$m. It was found that the relative error with respect to the TMM grows slightly with the number of monomers due to the overlapping of monomers along the line of sight. \citet{Tazaki2016} suggest that the relative errors should saturate for large $N$ because the degree of overlap is saturating for large $N$. With respect to TMM, the RGD theory tends to overestimate the scattered intensity, so it overestimates the geometric albedo of the particles.

From the equations above, we see that, for a given aggregate size, the scattered intensity increases with increasing fractal dimension (i.e., decreasing porosity).

\bsp    
\label{lastpage}
\end{document}